%% file: Halimi_Arxiv_UW.tex
\documentclass[12pt,journal,draftcls,a4paper,onecolumn]{IEEEtran}
\usepackage[nomarkers,nofiglist,notablist]{endfloat} % places all figures at the end of document
\usepackage{dsfont}
\usepackage{amsmath}
\usepackage{amssymb}
\usepackage{amsfonts}
\usepackage{graphicx}
\usepackage{amsmath}
\usepackage[all,poly]{xy}
\usepackage{multirow}
\usepackage{algorithm}
\usepackage{algorithmic}
\usepackage{color}
\usepackage[noadjust]{cite}
\usepackage{subfigure}
\usepackage{tikz,pgf}
\usetikzlibrary{arrows,automata}
\usepackage{verbatim}
\usetikzlibrary{positioning,shapes.geometric}
\usepackage{url}
\usepackage[T1]{fontenc}

\newcommand{\figwidth}{\columnwidth}
\newcommand{\beps}{\boldsymbol{\epsilon}}
\newcommand{\bthe}{\boldsymbol{\theta}}
\newcommand{\bThe}{\boldsymbol{\Theta}}

\input notations.tex

\input com_notations.tex
\title{Object Depth Profile and Reflectivity Restoration from Sparse Single-Photon Data Acquired in Underwater Environments}

\author{Abderrahim Halimi\thanks{The authors are with the School of Engineering and Physical Sciences, Heriot-Watt University, Edinburgh U.K.}, Aurora Maccarone, Aongus McCarthy,  Steve McLaughlin and Gerald S. Buller\thanks{This work was supported by the EPSRC Grants EP/J015180/1, EP/N003446/1, EP/M01326X/1, EP/K015338/1 and the DSTL National PhD Scheme. }} 

\begin{document}

%\ninept
%
\maketitle 
\begin{abstract}  
This paper presents two new algorithms for the joint restoration of depth and reflectivity (DR) images constructed from  time-correlated single-photon counting (TCSPC) measurements. Two extreme cases are considered: (i) a reduced acquisition time that leads to very low photon counts and (ii) a highly attenuating environment (such as a turbid medium) which makes the reflectivity estimation more difficult at increasing range.  Adopting a Bayesian approach, the Poisson distributed observations are combined with prior distributions about the parameters of interest, to build the joint posterior distribution. More precisely, two Markov random field (MRF) priors enforcing spatial correlations are assigned to the DR images. Under some justified assumptions, the restoration problem (regularized likelihood) reduces to a convex formulation with respect to each of the parameters of interest. This problem is first solved using an adaptive Markov chain Monte Carlo (MCMC) algorithm that approximates the minimum mean square parameter estimators. This algorithm is fully automatic since it adjusts the parameters of the MRFs by maximum marginal likelihood estimation. However, the MCMC-based algorithm exhibits a relatively long computational time. The second algorithm deals with this issue and is based on a coordinate descent algorithm. Results on single-photon depth data from laboratory based underwater measurements demonstrate the benefit of the proposed strategy that improves the quality of the estimated DR images.  
\end{abstract} 

%%%%%%%%%%%%%%%%%%%%%%%%%%%%%%%%%%%%%%%%%%%%%%%%%%%
%%%%%%%%%%%%%%%%%%%%%%%%%%%%%%%%%%%%%%%%%%%%%%%%%%%
%%%%%%%%%%%%%%%%%%%%%%%%%%%%%%%%%%%%%%%%%%%%%%%%%%%
\section{Introduction} \label{sec:Introduction} 
   
	%///// scenrio:  UW + single depth object + 
	%lidar-UW + Contrib1-formulation  + Contrib2-MCMC-ADMM 

Reconstruction of 3-dimensional scenes is a challenging problem encountered in many applications. For a given pixel, the time-of-flight light detection and ranging (Lidar) system achieves this goal by emitting laser pulses and recording the round-trip return time and intensity of the reflected signal \cite{AmannOE2001}. Single-photon Lidar typically uses a high repetition rate pulsed laser source in conjunction with a single-photon detector. The advantages of the single-photon approach are its shot-noise limited sensitivity, and its picosecond temporal response which can achieve millimeter-scale surface-to-surface resolution \cite{BullerJSTQE2007}. In single-photon Lidar, the recorded photon event is stored in a timing histogram which is formed by detecting photons from many laser pulses. The time delay and the amplitude of the histogram are related to the distance and reflectivity of the observed object, respectively, which allows the construction of the 3D scene. 

In this paper, we consider a scanning system whose acquisition time is defined by the user and is the same for each pixel, which leads to a deterministic and user-defined overall acquisition duration. Consequently, the number of detected photons can be larger than one for some pixels, whereas other pixels may be empty (i.e. no detected photons). We also assume solid target surfaces fabricated from opaque materials, so that only one reflection is observed in an individual pixel \cite{AltmannTIP2015a}. The study focuses on the following two extreme cases:  (i) a reduced data acquisition time and (ii) the use of an extremely attenuating medium \cite{MaccaroneOE2015}.  Both cases lead to a reduction in the number of detected photons per pixel, which affects the estimation of depth and target reflectivity. 
Indeed, taking underwater measurements leads to a severe attenuation of the intensity with respect to (w.r.t.) the target range, which makes the reflectivity estimation difficult. With such challenging scenarios, the measurement can be improved by, for example, increasing the laser power or the data acquisition time \cite{McCarthy_OE13,Kirmani_Science2014}, however this is not always practicable in a field situation. To use the available sparse photon data most efficiently, the alternative approach is to improve the processing of the acquired signals using signal processing techniques \cite{Wallace_Eurasip2010,AltmannTIP2015a,Dongeek_ICIP_2014,Halimi2016Eusipco_a}. The latter approach will be considered here to improve the estimated depth and reflectivity (DR) images for sparse single-photon data.

The first contribution of this paper is the use of a hierarchical Bayesian model associated with the DR images. Using the Poisson distribution of the observed photon counts, and introducing some approximations, lead to a log-concave likelihood distribution w.r.t. each of the parameters of interest. The resulting likelihood distribution is interesting for two reasons:  it allows the use of convex programming algorithms for parameter estimation and it is expressed w.r.t. preliminary estimates of the DR images which avoids the use of cumbersome photon count histograms during the refinement process.  Using Markov random fields (MRF), the parameters of interest are assigned prior distributions enforcing a spatial correlation between the pixels. More precisely, the depth image is  assigned an MRF distribution equivalent to a total variation (TV) prior \cite{RudinPD1992,BioucasDias2012}, while the reflectivity image is assigned a gamma-MRF prior  \cite{DikmenTASLP2010}. The likelihood and the prior distribution are then used to build the joint posterior distribution that is used for the parameter estimation.  
  
The second contribution of this paper is the derivation of two estimation algorithms associated with the proposed hierarchical Bayesian model. The first algorithm generates samples distributed according to the posterior using Markov chain Monte Carlo (MCMC) methods (such as the Gibbs sampler, and the Metropolis-Hastings algorithm) \cite{Robert1999}. These samples are then used to evaluate the  minimum-mean-square-error (MMSE) estimator of the DR images. This approach also allows the estimation of the regularization parameters, (the hyperparameters), associated with the MRF prior using the maximum marginal likelihood approach proposed in \cite{PereyraSSP2014}. Therefore, the MCMC method is fully automatic in the sense that it does not require the user to tune the model hyperparameters.  However, the resulting MCMC-based algorithm has a high computational complexity which can be a significant limitation for real time applications. The second algorithm deals with this limitation and approximates the maximum a posteriori (MAP) estimator by using a coordinate descent algorithm
\cite{Bertsekas1995,Sigurdsson2014}. The latter is used to sequentially update the different parameters to minimize the negative log-posterior, which is convex w.r.t. each parameter. In contrast to the reflectivity image that is updated analytically, the depth image is updated using the alternating direction method of multipliers (ADMM). This algorithm has shown good performance in different fields, both for the estimation quality and the reduced computational cost  \cite{Figueiredo_TIP2010,Afonso_TIP2011,Halimi2016Eusipco_a}.  
The proposed algorithms are complementary and represent useful tools to deal with different user requirements such as a reduced computational cost or an automatic hyperparameter estimation. 
Results on single-photon depth data acquired from laboratory experiments show the benefit of the proposed strategies that improve the quality of the estimated DR images. 

The paper is organized as follows. Section
\ref{sec:Observation_model}  introduces the observation model associated with the underwater photon counts. The proposed hierarchical Bayesian algorithm  for DR restoration is presented in Section  \ref{sec:Hierarchical_Bayesian_Model}. Section \ref{sec:Estimation algorithms}  introduces the two proposed estimation algorithms based on stochastic simulation and optimization.   Simulation results on synthetic data are reported in Section \ref{sec:Simulation_on_synthetic_data}.  Section \ref{sec:Simulation_on_real_data}  presents and analyzes results conducted using data acquired by an actual time-of-flight  scanning sensor based on TCSPC. Finally,  conclusions and future work are  reported in Section \ref{sec:Conclusions}.

%%%%%%%%%%%%%%%%%%%%%%%%%%%%%%%%%%%%%%%%%%%%%%%%%%%
%%%%%%%%%%%%%%%%%%%%%%%%%%%%%%%%%%%%%%%%%%%%%%%%%%%
%%%%%%%%%%%%%%%%%%%%%%%%%%%%%%%%%%%%%%%%%%%%%%%%%%%
%\section{Problem formulation} \label{sec:Problem_formulation} 

%%%%%%%%%%%%%%%%%%%%%%%%%%%%%%%%%%%%%%%%%%%%%%%%%%%
%%%%%%%%%%%%%%%%%%%%%%%%%%%%%%%%%%%%%%%%%%%%%%%%%%%
%%%%%%%%%%%%%%%%%%%%%%%%%%%%%%%%%%%%%%%%%%%%%%%%%%%
\section{Observation model} \label{sec:Observation_model} 
The Lidar observation $\bsy_{i,j,t}$ , where $(i,j) \in \left\lbrace1,\cdots,N_r\right\rbrace \times \left\lbrace 1,\cdots,N_c\right\rbrace$, represents the number of photon counts within the $t$th bin of the pixel  $(i,j)$. According to  \cite{HernandezPAMI2007,AltmannTIP2015a}, each photon count $\bsy_{i,j,t}$  is assumed to be drawn from the Poisson distribution $\mathcal{P} \left(.\right)$ as follows\vspace{-0.2cm}  
\begin{equation}
y_{i,j,t} \sim \mathcal{P} \left(s_{i,j,t}\right)  \vspace{-0.2cm}
\label{eqt:Statistical_model}
\end{equation} 
where $s_{i,j,t}$ is the average photon counts given by \cite{MaccaroneOE2015}
\begin{equation}
s_{i,j,t} = r_{i,j} e^{-\alpha t_{i,j}} g_0\left(t-t_{i,j} \right) + b_{i,j} 
\label{eqt:Observation_model}
\end{equation}
and $t_{i,j}\geq 0$ is the position of an object surface at a given range from the sensor (related to the depth),  $r_{i,j}\geq 0$ is the reflectivity of the target, $b_{i,j}\geq 0$ is a constant denoting the background and dark photon level, $\alpha$ represents the attenuation factor related to the transmission environment  and $g_0$ denotes the system impulse response assumed to be known from the calibration step. In air, the attenuation factor is $\alpha=0$ and the model \eqref{eqt:Observation_model} reduces to that studied in \cite{AltmannTIP2015a,Halimi2016Eusipco_a}. This paper considers the case of transmission under a highly attenuating environment in which $\alpha \geq 0$. In this case, the measured reflected intensity of the objects decreases as a function of
 their distance to the sensor which is valid for different scenarios such as highly scattering underwater measurements. Indeed, the single-photon depth images can be used underwater to localize objects such as boat wreckage, pipelines, etc.  The first objective of this paper is to estimate the target depth and reflectivity images of a target underwater or in any other extremely attenuating environment. 
The paper second objective deals with the extreme case of a very low photon counts per pixels. Under this scenario, it is possible to have missing pixels which have no received photons, i.e., $\sum_{t=1}^{T} y_{i,j,t}=0$. These missing pixels bring no information regarding the depth $t_{i,j}$ and reflectivity $r_{i,j}$ and should be considered separately from informative observed pixels as in \cite{CarlavanTIP2012}. %More precisely, we introduce a ($M\times N$) binary matrix $\bsK$ that contains a single non-zero value on each line to model the loss of some image pixels  
  %In this case, there will be some missing pixels (when no photons are present in a given pixels) that should be excluded from , the pixel can be considered 

%%%%%%%%%%%%%%%%%%%%%%%%%%%%%%%%%%%%%%%%%%%%%%%%%%%
%%%%%%%%%%%%%%%%%%%%%%%%%%%%%%%%%%%%%%%%%%%%%%%%%%%
%%%%%%%%%%%%%%%%%%%%%%%%%%%%%%%%%%%%%%%%%%%%%%%%%%% 
\section{Hierarchical Bayesian Model} \label{sec:Hierarchical_Bayesian_Model} 

This section introduces a hierarchical Bayesian model for estimating the target distance and reflectivity images of underwater measurements. The Bayesian approach accounts for both the statistical model associated with the observed data (likelihood) and the prior knowledge about the parameters of interest (prior distributions). This approach is interesting to alleviate the indeterminacy resulting from ill-posed problems and has been successfully applied to Lidar measurements in \cite{AltmannTIP2015a}. More precisely, if $f\left(\bThe \right)$  denotes the prior distribution assigned to the parameter $\bThe$, the Bayesian approach computes the posterior distribution of $\bThe$ using the Bayes rule
\begin{equation}
f(\bThe|\bsY)  \propto f(\bsY|\bThe) f(\bThe) \label{eqt:Bayes}
\end{equation} 
where $\propto$ means ``proportional to'' and $f(\bsY|\bThe)$ is the likelihood of the observation matrix $\bsY$ gathering all the observed pixels $y_{i,j,t}, \forall i,j,t$. The  MMSE and  MAP  estimators of $\bThe$ can be evaluated by the mean vector and maximum of this posterior. At this point, it is interesting to highlight the link between the Bayesian and optimization perspectives. Indeed, the MAP estimator can also be evaluated by minimizing the cost function obtained as the negative log-posterior function. From an optimization perspective, this cost function is considered as a regularized problem where the data fidelity term (likelihood) is constrained using some regularization terms (prior distributions). 
The following sections introduce the likelihood and the prior distributions (regularization terms) considered in this paper.   

\subsection{Likelihood} \label{subsec:Likelihood} 
Assuming independence between the observed pixels $y_{i,j,t}$ and considering the Poisson statistics leads to the following joint likelihood \vspace{-0.2cm}
\begin{equation}
P(\bsY | \bst,\bsr,\bsb) = \prod_{(i,j) \in \Omega} \prod_{t=1}^{T}{ \frac{s_{i,j,t}^{y_{i,j,t}}}{y_{i,j,t}!}  \exp^{-s_{i,j,t}}  } 
\label{eqt:Likelihood}
\end{equation}
where $\bst,\bsr,\bsb$ are $N\times 1$  vectors gathering the elements $t_{i,j},$ $ r_{i,j},$ $b_{i,j}, \forall i, \forall j$ (in lexicographic order), with $N= N_r N_c$,  $T$ is the total number of bins,  $\Omega$ gathers the indices of non-empty pixels
 and $s_{i,j,t}(\bst,\bsr,\bsb) $ has been denoted by $s_{i,j,t}$ for brevity.   
In a similar fashion to the classical estimation approach (see \cite{AltmannTIP2015a,Halimi2016Eusipco_a} for more details), this paper assumes the absence of the background level, i.e., $b_{i,j} =0$. Indeed, the underwater measurements are most often obtained in dark conditions (in the laboratory in our case) which justifies this assumption. Note, however, that the assumption is violated in presence of multiple scatterers, thus, its effect is studied when considering synthetic data. In addition to this simplification,  we further assume a Gaussian approximation for the instrument impulse response\footnote{The parameters $c_1$ and $\sigma^2$ can be estimated by fitting the actual impulse response with a Gaussian using a least squares algorithm.}  $g_0\left(t-t_{i,j} \right)= c_1  \exp^{- \frac{\left(t-t_{i,j}\right)^2}{2 \sigma^2}}$ as in \cite{AltmannTSP2015,Halimi2016Eusipco_a}, and that the temporal sum of the shifted impulse response  $c_2 =  \sum_{t=1}^{T}{g_0\left(t-t_{i,j} \right)}$ is a constant for all realistic target distances $t_{i,j}$ (which is justified when assuming that the observation time window is larger than the depth of the observed object).  
Under these assumptions, the likelihood reduces to $\mathcal{L} = \prod_{(i,j) \in \Omega}{\mathcal{L}_{i,j}}$  with (after removing unnecessary constants) 
\begin{equation}
 \mathcal{L}_{i,j} =  r_{i,j}^{c_2 r_{i,j}^{\textrm{ML0}}}  \exp^{\left[-\alpha c_2 r_{i,j}^{\textrm{ML0}}  t_{i,j} - \frac{\left(t_{i,j} - t_{i,j}^{\textrm{ML0}} \right)^2}{\frac{2 \sigma^2}{c_2 r_{i,j}^{\textrm{ML0}} } }  - c_2 r_{i,j}   \exp^{\left(-\alpha t_{i,j}\right)}  \right]} 
\label{eqt:likelihood}
\end{equation}
where $t_{i,j}^{\textrm{ML0}} = \frac{\left(\sum_{t=1}^{T}{t y_{i,j,t}}
\right)}{\left(\sum_{t=1}^{T}{y_{i,j,t}}
\right)}$ and  $r_{i,j}^{\textrm{ML0}} = \frac{1}{c_2}\left(\sum_{t=1}^{T}{y_{i,j,t}}
\right)$    are the maximum of this simplified likelihood w.r.t. $t_{i,j}$ and $r_{i,j}$ obtained in the air (with $\alpha=0$). The likelihood \eqref{eqt:likelihood} obtained is interesting for two reasons. First, it does not include the Lidar observation terms $y_{i,j,t}$ explicitly, which means that our formulation considers only the two observed images $r_{i,j}^{\textrm{ML0}}$ and  $t_{i,j}^{\textrm{ML0}}$ instead of the  $N_r\times N_c\times T$ matrix $y_{i,j,t}$. The computational cost is then drastically reduced when compared to the models studied in \cite{AltmannTIP2015a,Hernandez_MarinPAMI2008}   which considered the full $N_r\times N_c\times T$ data cube. Second, it is a log-concave distribution w.r.t. each of the parameters $t_{i,j}$ and $r_{i,j}$ separately,  that is suitable for the application of convex programming algorithms. 
%As mentioned previously, this paper deals with the extreme case of a very low photon counts per pixels. Under this scenario, it is possible to have missing pixels which have no photons, i.e., $s_{i,j}=0$. In addition, the Poisson distribution is not defined for a zero mean. These observations bring us to consider separately positive and null observed pixels as in \cite{CarlavanTIP2012}. More precisely, we introduce a ($M\times N$) binary matrix $\bsK$ that contains a single non-zero value on each line to model the loss of some image pixels  
  %In this case, there will be some missing pixels (when no photons are present in a given pixels) that should be excluded from , the pixel can be considered 
 Note finally that our approach can be interpreted as a joint depth-reflectivity image restoration problem of the estimates $t_{i,j}^{\textrm{ML0}}$ and $r_{i,j}^{\textrm{ML0}}$ that are of poor quality especially in the limit of very low photon counts or when acquiring the data in a significantly attenuating environment. The next section introduces the prior information introduced to improve the estimated images from \eqref{eqt:likelihood}.

\subsection{Priors for the distance image} \label{subsec:Priors_for_the_depth_image}
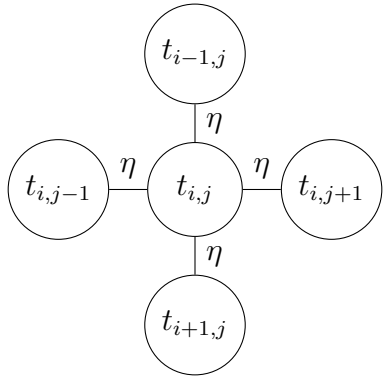
\begin{figure}[h!]
\centering 
\begin{tikzpicture}
 nodes %
\node[draw, circle, text centered] (center) {$t_{i+1,j}$}; 
\node[draw, circle,above =0.5  of center, text centered] (top1) {$~t_{i,j}~$};
\node[draw, circle,right =0.5  of top1, text centered] (tr) {$t_{i,j+1}$};
\node[draw, circle,left  =0.5  of top1, text centered] (tl) {$t_{i,j-1}$}; 
\node[draw, circle,above =0.5  of top1, text centered] (top2) {$t_{i-1,j}$}; 
%Lines
%\draw[-] (center) to  (w1r);
\draw[-] (center) -- (top1) node[right,midway] {$\eta$};
\draw[-] (top1) --(tr) node[above,midway] {$\eta$};
\draw[-] (top1) --(tl) node[above,midway] {$\eta$};
\draw[-] (top1) --(top2) node[right,midway] {$\eta$}; 
\end{tikzpicture} 
\caption{The total variation neighborhood structure. }\label{fig:TV_MRFs}
\end{figure}

The target distances exhibit correlation between adjacent pixels. This effect is accounted for by considering the following MRF prior distribution
\begin{equation}
f(\bst|\eta)  = \frac{1}{G(\eta)} \exp^{\left[-\eta \textrm{TV}(\bst)\right]}  \label{eqt:prior_parameters}
\end{equation}
where $G(\eta)$ is a normalizing constant,  $\eta$ is a coupling parameter that controls the amount of enforced spatial smoothness,  $\textrm{TV}(\bst) = \sum_{i,j}\sum_{(i',j')\in \upsilon(i,j)} |t_{i,j} - t_{i',j'}|$ denotes the total-variation regularization suitable for edge preservation \cite{RudinPD1992,BioucasDias2012} and $\upsilon(i,j)$ denotes the neighborhood of the pixel $(i,j)$ as shown in  Fig. \ref{fig:TV_MRFs}.

\subsection{Priors for the reflectivity image} \label{subsec:Priors_for_the_reflectivity_image}
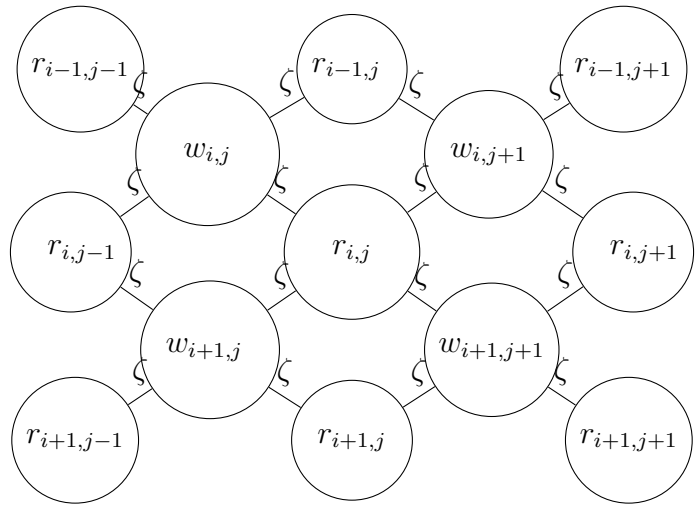
\begin{figure}[h!]
\centering 
\begin{tikzpicture}
 nodes %
\node[draw, circle, text centered] (center) {$~r_{i+1,j}~$}; 
\node[draw, circle,right =2  of center, text centered] (eps0r) {$r_{i+1,j+1}$};
\node[draw, circle,left  =2  of center, text centered] (eps0l) {$r_{i+1,j-1}$};
\node[above =0.25  of center, text centered] (top1) {$$};
\node[draw, circle,above =0.25  of top1, text centered] (top2) {$~~~r_{i,j}~~~$};
\node[above =0.25  of top2, text centered] (top3) {$$};
\node[draw, circle,above =0.25  of top3, text centered] (top4) {$r_{i-1,j}~$};

\node[draw, circle,right =0.8  of top1, text centered] (w1r) {$w_{i+1,j+1}$};
\node[draw, circle,left  =0.8  of top1, text centered] (w1l) {$~w_{i+1,j}~~$}; 

 \node[draw, circle,right =2  of top2, text centered] (eps2r) {$~~r_{i,j+1}$};
\node[draw, circle,left  =2  of top2, text centered] (eps2l) {$~~r_{i,j-1}$};

\node[draw, circle,left =0.8  of top3, text centered] (w3l) {$~~~w_{i,j}~~~$};
\node[draw, circle,right  =0.8  of top3, text centered] (w3r) {$~w_{i,j+1}~$};

 \node[draw, circle,right =2  of top4, text centered] (eps4r) {$r_{i-1,j+1}$};
\node[draw, circle,left  =2  of top4, text centered] (eps4l) {$r_{i-1,j-1}$}; 
%Lines
%\draw[-] (center) to  (w1r);
\draw[-] (center) -- (w1r) node[above,midway] {$\zeta$};
\draw[-] (center) --(w1l) node[above,midway] {$\zeta$};
\draw[-] (eps0r) --(w1r) node[above,midway] {$\zeta$};
\draw[-] (eps0l) --(w1l) node[above,midway] {$\zeta$};
\draw[-] (w1l) --(eps2l) node[above,midway] {$\zeta$};
\draw[-] (w1l) --(top2) node[above,midway] {$\zeta$};
\draw[-] (w1r) --(eps2r) node[above,midway] {$\zeta$};
\draw[-] (w1r) --(top2) node[above,midway] {$\zeta$};

\draw[-] (top2) -- (w3r) node[above,midway] {$\zeta$};
\draw[-] (top2) --(w3l) node[above,midway] {$\zeta$};
\draw[-] (eps2r) --(w3r) node[above,midway] {$\zeta$};
\draw[-] (eps2l) --(w3l) node[above,midway] {$\zeta$};

\draw[-] (w3l) --(eps4l) node[above,midway] {$\zeta$};
\draw[-] (w3l) --(top4) node[above,midway] {$\zeta$};
\draw[-] (w3r) --(eps4r) node[above,midway] {$\zeta$};
\draw[-] (w3r) --(top4) node[above,midway] {$\zeta$}; 
\end{tikzpicture}  
\caption{Gamma-MRF neighborhood structure. }\label{fig:GMRFs}
\end{figure}

Similarly as for the target distances, we expect the target reflectivity to vary smoothly from one pixel to another. This behavior is obtained by introducing an auxiliary variable $\bsw$ (of size  $N_{\textrm{r}}\times N_{\textrm{c}}$) and  assigning a gamma-MRF prior  for  $(\bsr,\bsw)$ as follows  \cite{DikmenTASLP2010,AltmannTCI2015b,HalimiTIP2016} 
\begin{eqnarray}
f\left(\bsw,\bsr | \zeta \right)  = &    \frac{1}{Z(\zeta)}  \prod_{(i,j)\in \nu_{\bsw}} {w_{i,j}^{-(4\zeta+1)}} \nonumber \\
\times & \prod_{(i',j')\in \nu_{\bsr}} {r_{i',j'}^{(4\zeta-1)}} \nonumber \\
\times & \prod_{((i,j),(i',j'))\in \mathcal{E} } {\exp\left( \frac{-\zeta r_{i',j'}}{w_{i,j}} \right)},
\label{eqt:priorEnergies}
\end{eqnarray}  
where $Z(\zeta)$ is a normalizing constant,  the partition  $\nu_{\bsw}$ (resp. $\nu_{\bsr}$) denotes the collection of variables $\bsw$ (resp. $\bsr$), the edge set $\mathcal{E}$ consists of pairs $(i,j)$ representing the connection between the variables and  $\zeta$ is a coupling parameter that controls the amount of spatial smoothness enforced by the GMRF. This prior ensures that each $r_{i,j}$ is connected to four neighbor elements of $\bsw$ and vice-versa (see Fig. \ref{fig:GMRFs}). The reflectivity coefficients $r_{i,j}$ are conditionally independent and the $1$st order neighbors (i.e., the spatial correlation) is only introduced via the auxiliary variables $\bsw$. An interesting property of this joint prior is that the conditional prior distributions of $\bsr$ and $\bsw$ reduce to conjugate inverse gamma ($\calI \calG$) and gamma ($\calG $) distributions as follows  
\begin{eqnarray}
w_{i,j} | \bsr,\zeta \sim   \calI \calG \left(4\zeta, 4\zeta \rho_{1,i,j}(\bsr) \right),
\nonumber \\
r_{i,j} | \beps,\zeta \sim     \calG \left(4\zeta, 1/(4\zeta \rho_{2,i,j}(\bsw))     \right),
\label{eqt:CondGam_IGam}
\end{eqnarray}
where  
\begin{eqnarray}
\rho_{1,i,j}(\bsr) = (r_{i,j} + r_{i-1,j} + r_{i,j-1} + r_{i-1,j-1})  /4,
 \nonumber \\
\rho_{2,i,j}(\bsw) = (w^{-1}_{i,j} + w^{-1}_{i+1,j} + w^{-1}_{i,j+1} + w^{-1}_{i+1,j+1})  /4.
\label{eqt:ParamGam_IGam}
\end{eqnarray}

\subsection{Posterior distribution} \label{subsec:Posterior_distribution}
\begin{figure}[h!]
\centering
\begin{tikzpicture}
 nodes %
\node[text centered] (Y) {$\bsY$};
\node[above =0.5  of Y, text centered] (top1) {};
\node[right =0.8  of top1, text centered] (T) {$\bst$};
\node[left =0.5  of top1, text centered] (RW) {$(\bsr,\bsw)$};
\node[above =0.5  of T, text centered] (top2T) {};
\node[draw, rectangle, right =0.5  of top2T, text centered] (tau1) {$\eta$};
\node[above =0.5  of RW, text centered] (top2RW) {};
\node[draw, rectangle, left =0.5  of top2RW, text centered] (tau2) {$\zeta$};
 edges %
\draw[->, line width= 1] (RW) to  [out=290,in=165, looseness=0.5] (Y); 
\draw[->, line width= 1] (T) to  [out=250,in=15, looseness=0.5] (Y);
\draw[->, line width= 1] (tau2) to  [out=290,in=150, looseness=0.5] (RW); 
\draw[->, line width= 1] (tau1) to  [out=290,in=15, looseness=0.5] (T);
\end{tikzpicture}
\caption{DAG for the parameter and hyperparameter priors. For the optimization algorithm, the user fixed hyperparameters appear in boxes. }
\label{fig:DAG}
\end{figure}
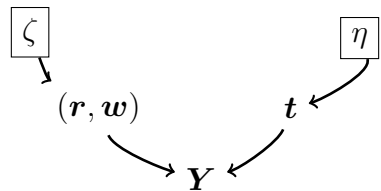

The  proposed Bayesian  model is illustrated by the directed acyclic graph (DAG) displayed in Fig. \ref{fig:DAG}, which highlights the relation between the observations $\bsY$, the parameters $\bst,\bsr,\bsw$ and the  hyperparameters $\eta,\zeta$. 
Assuming prior independence between the  parameter vector $\bThe = \left( \bst,\bsr,\bsw \right)$, the joint posterior distribution associated with the proposed Bayesian model is given by
\begin{equation}
f\left(\bThe | \bsY,\eta,\zeta \right)  \propto f(\bsY|\bThe)  f\left(\bThe| \eta,\zeta \right).
\label{eqt:Joint_Posterior}
\end{equation}
This posterior will be used to evaluate the Bayesian estimators of $\bThe$. For this purpose, we propose two algorithms based on an MCMC and an optimization approach. The first approach uses an MCMC approach to evaluate the MMSE estimator of $\bThe$ by generating samples according to the joint posterior distribution. Moreover, it allows the estimation of the hyperparameters  $\eta,\zeta$ by using a maximum marginal likelihood estimation during the inference procedure (as detailed in the next section). However, this MCMC algorithm presents a significant computational complexity which can limit the applicability for real time applications. The second optimization algorithm deals with this issue and provides fast MAP estimates for $\bThe$. This is achieved by maximizing the posterior \eqref{eqt:Joint_Posterior} w.r.t. $\bThe$,  or equivalently, by minimizing the negative log-posterior given by $ \mathcal{F} = - \textrm{log} [ f\left(\bThe | \bsY,\eta,\zeta \right)]$. Note however, that the hyperparameters are fixed under this approach. The two estimation algorithms are described in the next section.

%%%%%%%%%%%%%%%%%%%%%%%%%%%%%%%%%%%%%%%%%%%%%%%%%%%
%%%%%%%%%%%%%%%%%%%%%%%%%%%%%%%%%%%%%%%%%%%%%%%%%%%
%%%%%%%%%%%%%%%%%%%%%%%%%%%%%%%%%%%%%%%%%%%%%%%%%%%
\section{Estimation algorithms} \label{sec:Estimation algorithms} 

%%%%%%%%%%%%%%%%%%%%%%%%%%%%%%%%%%%%%%%%%%%%%%%%%%%
\subsection{MCMC algorithm} \label{subsec:MCMC_algorithm} 
%///// highlight the main diff. with Yoann's algo.

The principle of the MCMC approach is to generate samples whose stationary distribution is the desired posterior distribution \eqref{eqt:Joint_Posterior}. The distribution \eqref{eqt:Joint_Posterior} being difficult to sample,  the Gibbs algorithm can be used to iteratively generate samples according to its conditional distributions  \cite{Robert1999}. Moreover,  when a conditional distribution cannot be sampled directly, sampling techniques such as the Metropolis-Hasting (MH) algorithm can be applied leading to a Metropolis-within-Gibbs sampler.   In this paper, we  generate samples associated with the parameters $\left(\bthe_{1},\bthe_{2},\bthe_{3} \right)=\left( \bst,\bsr,\bsw \right)$ and use them to approximate the MMSE estimators given by
\begin{equation}
\hat{\bthe_{i}}^{\textrm{MMSE}} =   \mathds{E} \left[\bthe_{i} | \bsY, \hat{\eta}, \hat{\zeta}\right], \textrm{ for  } i=1,2,3  
\label{eqt:MMSE_est}
\end{equation}
where the expectation $\mathds{E}(.)$ is taken w.r.t. the marginal posterior density $f \left(\bthe_{i} | \bsY, \hat{\eta}, \hat{\zeta}\right)$  (by marginalizing $\bthe_{j}, j\neq i$, this density takes into account their uncertainty). In addition to these parameters, the hyperparameters $\eta, \zeta$ are also estimated by considering the method proposed in \cite{PereyraSSP2014}, which is based on the maximum marginal likelihood estimator, given by 
\begin{equation}
\left(\hat{\eta}, \hat{\zeta} \right) =  \operatornamewithlimits{\textrm{argmax}}\limits_{\eta \in \mathds{R}^+, \zeta \in \mathds{R}^+}   f \left(\bsY| \eta, \zeta\right). 
\label{eqt:MMAP}
\end{equation}
This method provides a point estimate for the hyperparameters that is used to evaluate the parameter MMSE as indicated in \eqref{eqt:MMSE_est}. These approaches have two main advantages: (i) it allows for an automatic adjustment of the value of $\left(\eta, \zeta\right)$ for each image which leads to an estimation improvement, (ii) it has a reduced computational cost when compared to competing approaches \cite{Pereyra2013b}. It should be noted that the resulting algorithm is similar to \cite{AltmannTIP2015a}  while the main differences relate to the different estimated parameters, the different distribution expressions (due to the underwater observation model), and to the discrete target positions  in \cite{AltmannTIP2015a} which are continuous in this paper.
The next subsections provide more details regarding the main steps of the sampling algorithm.

\subsubsection{Sampling the target positions} \label{subsubsec:target_positions} 
The conditional distribution of $\bst$ is given by 
\begin{equation}
f\left(\bst | \bsr, \bsY\right)    \propto   \exp^{- \mathcal{C}(\bst)}, \label{eqt:cond_T}
\end{equation}
with
\begin{eqnarray}
\mathcal{C}(\bst) &=&  \sum_{(i,j) \in \Omega}{\left[\frac{\left(t_{i,j} - t_{i,j}^{\textrm{ML0}} + \alpha \sigma^2  \right)^2}{\frac{2 \sigma^2}{c_2 r_{i,j}^{\textrm{ML0}} } }  + c_2 r_{i,j}   \exp^{\left(-\alpha t_{i,j}\right)}\right]} \nonumber \\
&+& \textit{i}_{\mathds{R}_{+}}\left(\bst\right) + \eta  \textrm{TV}\left( \bst\right)   
\label{eqt:Cost_funcT}
\end{eqnarray} 
%\begin{eqnarray}
%f(t_{i,j} | r_{i,j}, \bsy_{i,j}) & \propto & \exp^{\left[ - \frac{\left(t_{i,j} - t_{i,j}^{\textrm{ML0}} + \alpha \sigma^2  \right)^2}{\frac{2 \sigma^2}{c_2 r_{i,j}^{\textrm{ML0}} } }  - c_2 r_{i,j}   \exp^{\left(-\alpha t_{i,j}\right)}  \right]}  \nonumber\\
%& \times & \exp^{- \eta  \sum_{(i',j')\in \upsilon(i,j)} |t_{i,j} - t_{i',j'}|}. 
%\label{eqt:cond_T}
%\end{eqnarray}
%for $(i,j) \in \left\lbrace1,\cdots,N_r\right\rbrace \times \left\lbrace 1,\cdots,N_c\right\rbrace$, where $f\left(\bst | \bsr, \bsY\right) = \prod_{i,j} f(t_{i,j} | r_{i,j}, \bsy_{i,j})$ and
where  the observations $\bsY$ are introduced via the images $r_{i,j}^{\textrm{ML0}}$ and $t_{i,j}^{\textrm{ML0}}$ and  $\textit{i}_{\mathds{R}_{+}}\left(\bst\right)$ is the nonnegative orthant indicator function. Since it is not easy to sample according to \eqref{eqt:cond_T}, we propose to update the target positions using a Metropolis-Hasting (MH) move. More precisely, a new position is proposed following a Gaussian random walk procedure (the variance of the proposal distribution has been adjusted to obtain an acceptance rate close to $0.5$, as recommended in \cite{Robertmcmc}). Note finally that the independent positions (positions that are not directly related by the MRF-TV structure) are sampled in parallel using a check-board scheme, which accelerates the sampling procedure. 

\subsubsection{Sampling the reflectivity coefficients} \label{subsubsec:reflectivity} 
Using \eqref{eqt:likelihood} and \eqref{eqt:priorEnergies}, it can be easily shown that $\bsr$, and  $\bsw$ are distributed according to the following gamma and inverse gamma distributions 
\begin{eqnarray}
r_{i,j} | \beps,\zeta & \sim &    \calG \left(4\zeta+ c_2 k_{i,j} r_{i,j}^{\textrm{ML0}}, \frac{1}{\beta_{i,j}}     \right), \label{eqt:posterior_R}  \\
w_{i,j} | \bsr,\zeta & \sim &  \calI \calG \left(4\zeta, 4\zeta \rho_{1,i,j}(\bsr) \right),
\label{eqt:posterior_W}
\end{eqnarray}
where $\beta_{i,j} = 4\zeta \rho_{2,i,j}(\bsw)+ c_2 k_{i,j} \exp^{(-\alpha t_{i,j})}$, $k_{i,j}=0$ if the pixel is missing and $k_{i,j}=1$ otherwise (non-empty observed pixel). As a consequence, sampling according to \eqref{eqt:posterior_R} and \eqref{eqt:posterior_W} is straightforward.

\subsubsection{Updating the MRF parameters} \label{subsubsec:Updating_the_MRF_parameters} 
The MRF parameters maximizing the marginal likelihood $ f \left(\bsY| \eta, \zeta\right)$ are updated using the approach proposed in  \cite{PereyraSSP2014}. As reported in \cite{PereyraSSP2014,AltmannTIP2015a}, this approach provides a good approximation of the MRF parameters while requiring a reduced computational cost when compared to alternative approaches \cite{Pereyra2013b}. At each iteration of the MCMC algorithm, $\eta$ and $\zeta$ are updated as follows

{\small{\begin{eqnarray}
\eta^{(n+1)} & = & \mathcal{P}_{[0,\eta_{\textrm{max}}]}\left\lbrace\eta^{(n)} +  \varsigma_n  \left[TV\left(\bst^{(n)}\right) -  TV\left(\bst'\right) \right] \right\rbrace  \label{eqt:gradient_eta}   \\
\zeta^{(n+1)} & = & \mathcal{P}_{[0,\zeta_{\textrm{max}}]}\left\lbrace \zeta^{(n)} +  \varsigma_n  \left[\phi\left(\bsr^{(n)},\bsw^{(n)}\right) -  \phi\left(\bsr',\bsw'\right) \right]   \right\rbrace \nonumber\\
\label{eqt:gradient_zeta} 
\end{eqnarray}}}
\noindent where $\varsigma_n = n^{-3/4}$, $\mathcal{P}_{[a,b]} (x)$ denotes the projection operator of $x$ in the interval $[a,b]$,  and $\phi\left(\bsr,\bsw\right) = - 4 \sum_{(i,j)\in \nu_{\bsw}} {\log\left(w_{i,j}\right)}  + 4 \sum_{(i',j')\in \nu_{\bsr}} {\log\left(r_{i',j'}\right)} 
- \sum_{((i,j),(i',j'))\in \mathcal{E} } {\left( \frac{r_{i',j'}}{w_{i,j}} \right)}.$  These expressions originate from a projected gradient descent step in which the intractable gradients $\frac{\partial}{\partial \eta} \log{f \left(\bsY| \eta^{(n)}, \zeta^{(n)}\right)}$ and $\frac{\partial}{\partial \zeta} \log{f \left(\bsY| \eta^{(n)}, \zeta^{(n)}\right)}$ have been approximated by the biased estimators $\left[TV\left(\bst^{(n)}\right) -  TV\left(\bst'\right) \right]$ and  $\left[\phi\left(\bsr^{(n)},\bsw^{(n)}\right) -  \phi\left(\bsr',\bsw'\right) \right]$. These estimators use  the current samples $\bst^{(n)}$, $\bsr^{(n)}$, $\bsw^{(n)}$ and other auxiliary samples  $\bst',\bsr',\bsw' $ generated with kernels $\mathcal{K}_1$ and $\mathcal{K}_2$ whose target distributions are \eqref{eqt:prior_parameters} and \eqref{eqt:priorEnergies}, respectively (see Algo. \ref{alg:MCMC_algorithm}). Note also that the values obtained are projected using the operator $ \mathcal{P}$ to guarantee the positivity constraints of $\eta$ and $\zeta$  and the stability of the stochastic optimization algorithm ($\eta_{\textrm{max}} = \zeta_{\textrm{max}}=20$ in the following). Note finally that the hyperparameters are only updated in the burn-in period ($n < N_{\textrm{bi}}$)  and are fixed to their final values for the useful samples $N_{\textrm{bi}} \leq n \leq  N_{\textrm{MC}}$.
 Finally we refer the reader to  \cite{PereyraSSP2014,AltmannTIP2015a} for more details regarding this procedure.

\begin{algorithm}
\caption{MCMC algorithm} \label{alg:MCMC_algorithm}
\begin{algorithmic}[1]
\STATE \underline{Input} $N_{\textrm{bi}}$, $N_{\textrm{MC}}$ and the impulse response parameters $c_1, \sigma^2$
       \STATE \underline{Initialization}
       \STATE Initialize parameters $\bst^{(0)},\bsr^{(0)},\bsw^{(0)},\eta^{(0)}$, and $\zeta^{(0)}$  
       \STATE \underline{Update parameters/hyperparameters}
       \FOR{$n=1:N_{\textrm{MC}}$}
               \STATE Sample $\bst^{(n)}$  according to \eqref{eqt:cond_T} using MH
               \STATE Sample $\bsr^{(n)}$ according to  \eqref{eqt:posterior_R} 
               \STATE Sample $\bsw^{(n)}$ according to  \eqref{eqt:posterior_W}
							\IF{$n < N_{\textrm{bi}}$}  
							 \STATE Sample $\bst' \sim \mathcal{K}_1(\bst | \bst^{(n)}, \eta^{(n-1)})$   
               \STATE Sample $(\bsr',\bsw') \sim \mathcal{K}_2(\bsr,\bsw | \bsr^{(n)},\bsw^{(n)}, \zeta^{(n-1)})$  
							 \STATE Update $\eta$  using \eqref{eqt:gradient_eta} 
               \STATE Update $\zeta$  using \eqref{eqt:gradient_zeta} 
							\ENDIF
       \ENDFOR
			\STATE \underline{Output}  $\left\lbrace\bst^{(n)},\bsr^{(n)}\right\rbrace_{n=1}^{N_{\textrm{MC}}}$
\end{algorithmic}
\end{algorithm}

%%%%%%%%%%%%%%%%%%%%%%%%%%%%%%%%%%%%%%%%%%%%%%%%%%%
%%%%%%%%%%%%%%%%%%%%%%%%%%%%%%%%%%%%%%%%%%%%%%%%%%% 
\subsection{Optimization algorithm} \label{subsec:Optimization_algorithm} 

This section describes an alternative to the MCMC algorithm which is based on a fast optimization algorithm.  The latter maximizes the joint posterior  \eqref{eqt:Joint_Posterior}  w.r.t. the parameters of interest to approximate the MAP estimator of $\bThe$. The resulting optimization problem is tackled using a coordinate descent algorithm (CDA) \cite{Bertsekas1995,Sigurdsson2014,HalimiTGRS2015} that sequentially updates the different parameters as illustrated in Algo. \ref{alg:Coordinate_descent_algorithm}.  Thus, the algorithm iteratively updates each parameter by maximizing its conditional distribution as described in the following subsections.

\subsubsection{Updating the target positions} \label{subsubsec:Updating_the_target_positions} 
Maximizing the conditional distribution of the target positions \eqref{eqt:cond_T} is equivalent to minimizing its negative logarithm $\mathcal{C}(\bst)$, given by \eqref{eqt:Cost_funcT}. The latter is a proper, lower semi-continuous, coercive and strictly convex (since $r_{i,j}^{\textrm{ML0}}>0,  r_{i,j}>0$) function w.r.t. $\bst$, so that there exists a unique minimizer of $\mathcal{C}(\bst)$ (see the Appendix).% for example \cite{Bertsekas1995,CombettesMMS2005,Figueiredo_TIP2010}). 
This problem can be solved using many convex programing algorithms \cite{CombettesIP2008,Figueiredo_TIP2010,Afonso_TIP2011,Boyd2004_CP}. In this paper, we consider the ADMM variant proposed in  \cite{BioucasDias2012} that has shown good performance in many fields \cite{BioucasWhispers2010,Halimi2016Eusipco_a}  while requiring a reduced computational cost. This algorithm is theoretically ensured to reach the unique minimum of $\mathcal{C}(\bst)$. More details regarding this algorithm and its convergence properties are provided in the Appendix.

%%% quad:  str. convex (for rML0>0), proper, coercive, lsc
%%% expo:  str. convex (for r_ij>0), proper,  lsc
%%% quad+expo+iR+: str. convex (for rML0>0,r_ij>0), proper, coercive, lsc
%%% TV:  convex, proper,  lsc
%%% quad+expo+iR+: str. convex (for rML0>0,r_ij>0), proper, coercive, lsc

\subsubsection{Updating the reflectivity coefficients} \label{subsubsec:Updating_the_reflectivity_coefficients}  
Similarly to the target positions, maximizing the conditional distribution of  $\bsr$ (resp.  $\bsw$)  provided in \eqref{eqt:posterior_R} (resp.  \eqref{eqt:posterior_W}) is equivalent to minimizing $\mathcal{C}_1$ (resp. $\mathcal{C}_2$) given by
\begin{eqnarray} 
\mathcal{C}_1 (\bsr) & = &   \sum_{i,j} { (1-4\zeta- c_2 k_{i,j} r_{i,j}^{\textrm{ML0}}) \log(r_{i,j}) + \frac{r_{i,j}}{\beta_{i,j}} } \label{eqt:Prob_R}\\ 
\mathcal{C}_2 (\bsw) & = &  \sum_{i,j} { (4\zeta+1)  \log(w_{i,j})  + \frac{4\zeta \rho_{1,i,j}(\bsr)}{w_{i,j}}  }. \label{eqt:Prob_W} 
\end{eqnarray}
The minimum of these functions is uniquely attained and given by 
\begin{eqnarray} 
\overline{r_{i,j}} & = & \frac{4\zeta+ c_2 k_{i,j} r_{i,j}^{\textrm{ML0}}-1}{\beta_{i,j}}, \forall i,j  \label{eqt:mode_R}\\  
\overline{w_{i,j} } & = & \frac{4\zeta \rho_{1,i,j}(\bsr)}{4\zeta+1}, \forall i,j    \label{eqt:mode_W} 
\end{eqnarray}
subject to $4\zeta+ c_2 r_{i,j}^{\textrm{ML0}}>1$ which is always satisfied for $\zeta>0.25$.
These solutions are used to update the parameters  $\bsr$ and $\bsw$ as shown in Algo. \ref{alg:Coordinate_descent_algorithm}.

\subsubsection{Convergence and stopping criteria} \label{subsubsec:Convergence_and_stopping_criteria} 
The proposition 2.7.1  in  \cite{Bertsekas1995} asserts that the limit points of the sequence generated by the coordinate descent algorithm  ($\bThe^{n}$ for the $n$th iteration) are stationary points of $ \mathcal{F} = - \textrm{log} [ f\left(\bThe | \bsY,\eta,\zeta \right)]$  provided that the minimum of that function w.r.t. $\bThe$ along each coordinate is unique and that the function $ \mathcal{F} $ is monotonically non-increasing along each coordinate in the interval from $\bthe^{n}_{i}$ to $\bthe^{n+1}_{i}$. These conditions are satisfied for the parameters considered.  Indeed, the estimation of the target positions is a convex minimization problem whose solution is uniquely attained by the ADMM algorithm. Along the  reflectivity coordinate, the function $\mathcal{C}_1$ is convex and has a unique minimum (for $\zeta>0.25$). Along the  auxiliary variable coordinate, $\mathcal{C}_2$  has a unique minimum and is monotonically non-increasing on each side of the minimum. These satisfy the conditions of the proposition 2.7.1  in  \cite{Bertsekas1995}.
Moreover, note  that the cost function $ \mathcal{F}$ is not convex, thus, the solution obtained might depend on the initial values that should be chosen carefully. Therefore, the reflectivity and target positions are initialized using  the result of the classical  approach (known as X-corr algorithm \cite{AltmannTIP2015a}). For each pixel, this approach estimates the reflectivity by $r_{i,j}^{\textrm{ML0}}$ and the depth by finding the maximum of the cross-correlation of the histogram $\bsy_{i,j}$ with the impulse response $g_0$  (see \cite{AltmannTIP2015a} for more details regarding the X-corr algorithm).  With these initializations, the proposed algorithm reached minima of ``good quality'' in the considered simulations (see Sections \ref{sec:Simulation_on_synthetic_data}  and \ref{sec:Simulation_on_real_data}).

Two stopping criteria have been considered for Algo. \ref{alg:Coordinate_descent_algorithm}.  The first criterion compares the new value of the cost function to the previous one and stops the algorithm if the relative error between these two values is smaller than a given threshold, i.e.,
\begin{equation}
| \mathcal{F} \left(\bThe^{t+1}\right)-\mathcal{F} \left(\bThe^{t}\right) | \leq  \delta \mathcal{F} \left(\bThe^{t}\right),
\label{eqt:criteria1}
\end{equation}
where $|.|$ denotes the absolute value. 
The second criterion is based on a maximum number of iterations $N_{\textrm{max}}$. These values  have been fixed empirically to  $(\delta,N_{\textrm{max}}) = (10^{-2},500)$ in the rest of the paper.

\begin{algorithm}
\caption{Coordinate descent algorithm (CDA)} \label{alg:Coordinate_descent_algorithm}
\begin{algorithmic}[1]
\STATE \underline{Input}  $N_{\textrm{max}}, c_1, \sigma^2, \eta, \zeta$
       \STATE \underline{Initialization}
       \STATE Initialize parameters $\bst^{(0)},\bsr^{(0)},\bsw^{(0)}$  and $n \leftarrow 1$ 
       \STATE conv$ \leftarrow 0$,
       \STATE \underline{Parameter update}
       \WHILE{conv$=0$}
			         \STATE Update $\bst^{(n)}$  using Algo. \ref{alg:ADMM_for_depth_estimation}
               \STATE Update $\bsr^{(n)}$ according to \eqref{eqt:mode_R}
               \STATE Update $\bsw^{(n)}$ according to \eqref{eqt:mode_W}
               \STATE Set conv$ \leftarrow 1$ if the convergence criteria are satisfied
               \STATE $n  \leftarrow  n + 1$
       \ENDWHILE 
\end{algorithmic}
\end{algorithm}

%%%%%%%%%%%%%%%%%%%%%%%%%%%%%%%%%%%%%%%%%%%%%%%%%%%
%%%%%%%%%%%%%%%%%%%%%%%%%%%%%%%%%%%%%%%%%%%%%%%%%%%
%%%%%%%%%%%%%%%%%%%%%%%%%%%%%%%%%%%%%%%%%%%%%%%%%%%

\section{Simulation on synthetic data} \label{sec:Simulation_on_synthetic_data} 
This section evaluates the performance of the proposed algorithms on synthetic data with a known ground truth. All simulations have been implemented using MATLAB R2015a on a computer with Intel(R) Core(TM) i7-
4790 CPU@3.60GHz and 32GB RAM. The section is divided into two parts whose objectives are: 1) introducing the criteria used for the evaluation of the estimation results,  and 2) analysis of the algorithms performance for different background levels. 

%%%%%%%%%%%%%%%%%%%%%%%%%%%%%%%%%%%%%%%%%%%%%%%%%%%
%%%%%%%%%%%%%%%%%%%%%%%%%%%%%%%%%%%%%%%%%%%%%%%%%%%
\subsection{Evaluation criteria} \label{subsec:Evaluation_criteria} 
The restoration quality was evaluated qualitatively by visual inspection and quantitatively using the  signal-to-reconstruction error ratio, $\textrm{SRE} = 10 \log_{10} \left(\frac{||\bsx||^2}{ ||\bsx- \widehat{\bsx}||^2  } \right)$, where $\bsx$ is the reference depth or reflectivity image ,  $\widehat{\bsx}$ is the restored image and $||\bsx||^2$ denotes the $\ell_2$ norm given by $\bsx^T \bsx$. The returned values of this criterion are in decibel, the higher the better.  The reference images are known for synthetic images. For real data,  the estimated images with the MCMC approach in clear water, and with the highest acquisition time are considered as reference maps. As a result of the assumption of the absence of background photons, the proposed algorithms may be biased in a highly scattering environment. This effect is evaluated by considering the normalized-bias criterion given by $\textrm{N-Bias} = \frac{|\mathds{E}[\bsx- \widehat{\bsx}]|}{|\mathds{E}[\bsx]|}$.

We also provide some measures that are used in the experimental sections. We define one attenuation length (AL) as the distance after which the transmitted light power is reduced to $1/e$ of its initial value. If a target is located at range $d$ from the sensor, its stand-off distance expressed in AL can be computed as $\textrm{AL} = \alpha d$.  This measure is commonly used to highlight the attenuation affecting a given target \cite{MaccaroneOE2015}, and will be considered when processing real data. Similarly to \cite{PellegriniMST2000}, we consider two other measures related to the background level. The first is the signal-to-background ratio given by   $\textrm{SBR} = \frac{r c_1}{b}$.  The second is the signal-to-noise ratio given by $\textrm{SNR} = \frac{r c_1}{\sqrt{r c_1+b}}$.
% Note that both criteria are increasing function with $r$ and that their maxima (denoted by $\textrm{SBR}_m$ and $\textrm{SNR}_m$) are attained for $r=1$.

%%%%%%%%%%%%%%%%%%%%%%%%%%%%%%%%%%%%%%%%%%%%%%%%%%%
%%%%%%%%%%%%%%%%%%%%%%%%%%%%%%%%%%%%%%%%%%%%%%%%%%%
\subsection{Effect of the background} \label{subsec:Effect_of_the_background}  
\begin{figure}[h!]
\centering
\includegraphics[width=1.05\figwidth]{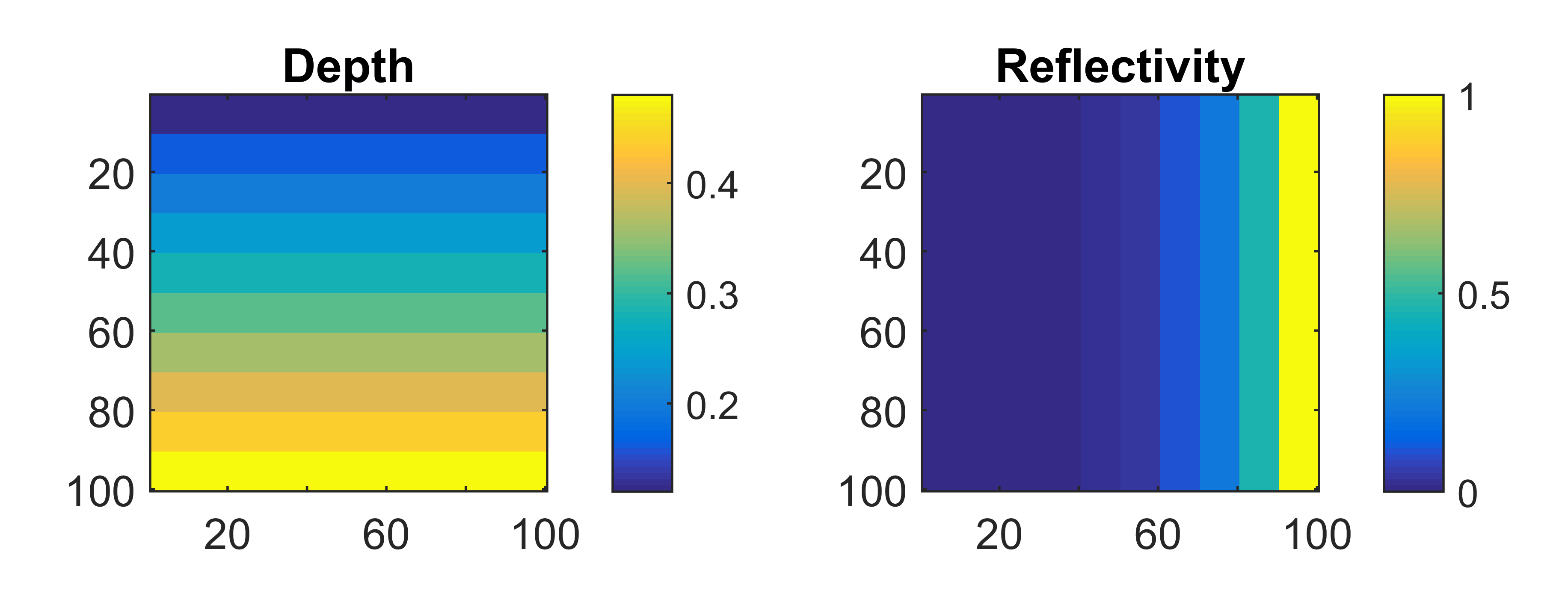}
\caption{Synthetic depth and reflectivity images.} \label{fig:Results_SynthB_Images}
\end{figure}

\begin{figure}[h!]
\centering
\includegraphics[width=1.05\figwidth]{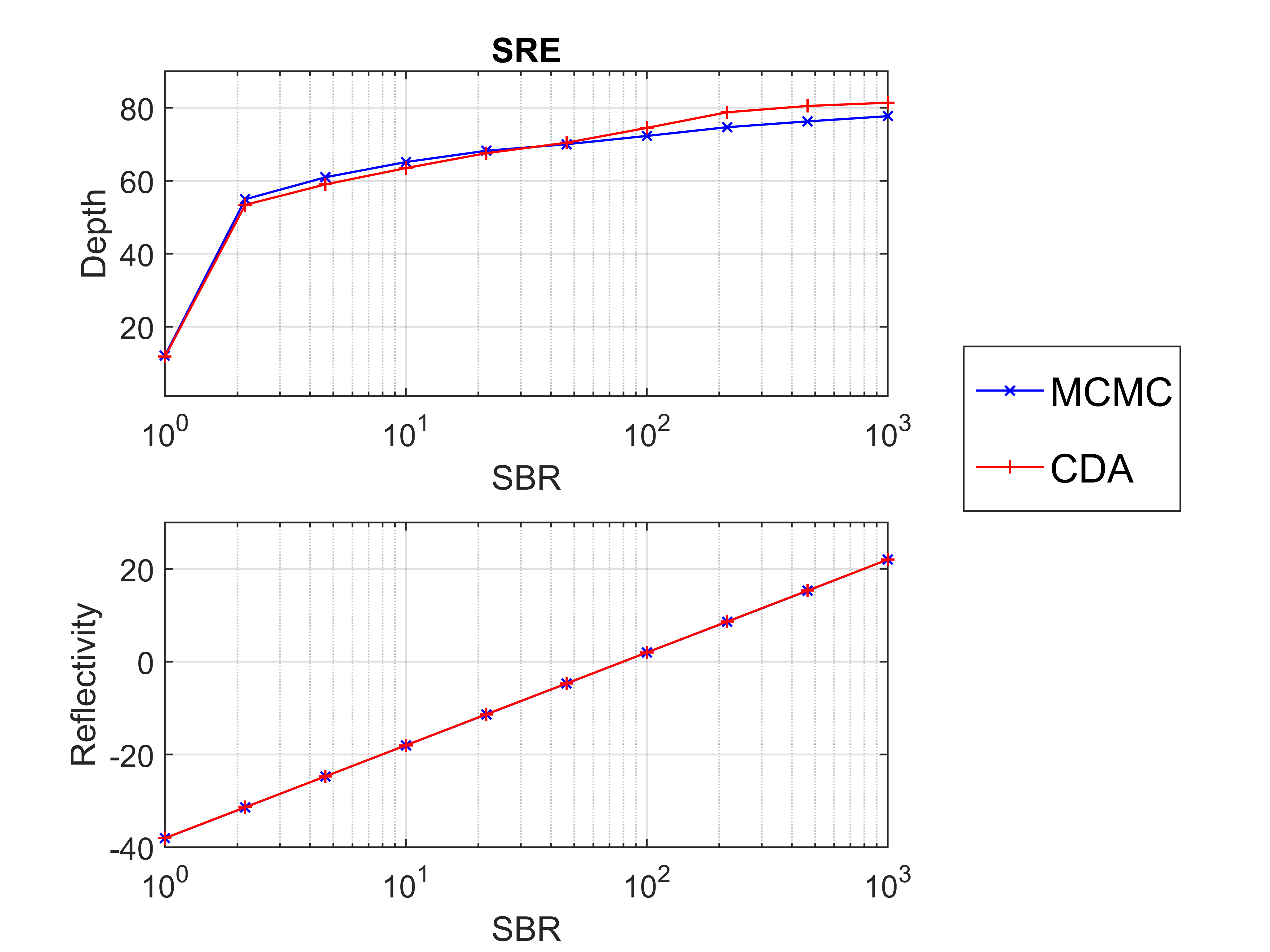}
\caption{SRE of depth and reflectivity with respect to the background levels for the MCMC (in blue) and CDA  (in red) algorithms.} \label{fig:Results_SynthB_SRE}
\end{figure}

\begin{figure}[h!]
\centering
\includegraphics[width=1.05\figwidth]{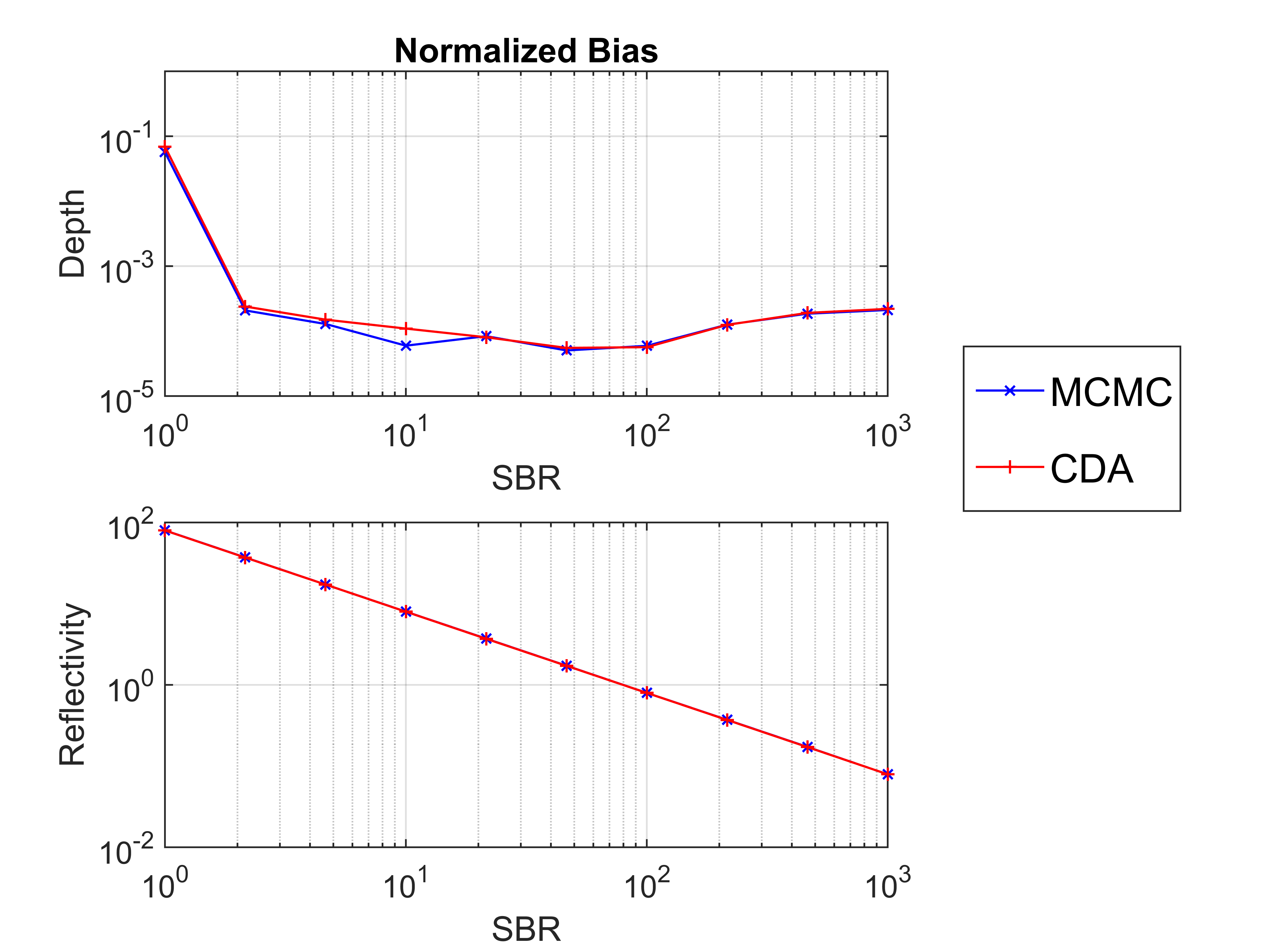}
\caption{Normalized bias of depth and reflectivity with respect to the background levels for the MCMC (in blue) and CDA  (in red) algorithms.} \label{fig:Results_SynthB_NBIAS}
\end{figure}

In a highly scattering environment or with reduced acquisition times, the background level might increase w.r.t. the useful signal. This section evaluates this effect when considering synthetic (computer-simulated)  data. A synthetic data cube has been generated according to model  \eqref{eqt:Statistical_model} with the following parameters $\alpha=0, c_1 = 1000, \sigma^2 = 100, b_{i,j}=1, \forall i,j$, $N_r=100$ pixels, $N_c=100$ pixels, and $T= 2000$ time bins  where a time bin represents $2$ picoseconds. The depth distance $d$ corresponding to $T$ bins can be computed as follows $d = \frac{T c}{2 n_e}$, where $c$ is the speed of light and $n_e$ is the refractive index of the propagation environment ($n_e=1$ for the air and $n_e=1.33$ for water). The synthetic data contains ten depths in the range $ [12, 48]$ cm  and ten reflectivity levels in the interval $r_{i,j} \in [0,1]$, as shown in Fig. \ref{fig:Results_SynthB_Images}.   The DR images are estimated using the proposed MCMC and CDA algorithms. The CDA algorithm requires the regularization parameters to be set manually. In this study, we provide the best performance (in terms of SRE) of this algorithm when testing the following values  $\eta  \in \left[0.01, 0.1, 0.5, 1, 2, 5\right]$ and  $\zeta  \in \left[0.3,5,10\right]$.   The performance analysis is conduced w.r.t. the SBR criterion that evaluates the ratio between the useful signal levels  $r_{i,j} c_1$ (whose variation depend on the reflectivity levels shown in Fig. \ref{fig:Results_SynthB_Images}) and the background levels $b_{i,j}=1, \forall i,j$. Fig. \ref{fig:Results_SynthB_SRE} shows the obtained SRE for depth and reflectivity w.r.t. SBR. Overall, the proposed algorithms provide similar performance. For both depth and reflectivity, the figure shows a decreasing performance when the SBR ratio decreases. However, the depth SRE remains high even for  $\textrm{SBR}=1$. The reflectivity performance decreases log-linearly w.r.t. the SBR ratio and attains low SRE values for $\textrm{SBR}=1$. This is mainly due to a reflectivity estimation bias in the presence of a high background level. Fig. \ref{fig:Results_SynthB_NBIAS} highlights this behavior and shows the estimation bias for depth and reflectivity. While the depth bias is always lower than $10 \%$,  the reflectivity shows high biases for low  $\textrm{SBR}=1$ which explains the low SRE values. This bias can be corrected when processing real data  using a look-up-table, however, this is beyond the scope of this paper.
These results highlight the sensitivity of the estimated reflectivity to the background level while they confirm the good estimation of the depth image even for low SBR.

%%%%%%%%%%%%%%%%%%%%%%%%%%%%%%%%%%%%%%%%%%%%%%%%%%%
%%%%%%%%%%%%%%%%%%%%%%%%%%%%%%%%%%%%%%%%%%%%%%%%%%%
%%%%%%%%%%%%%%%%%%%%%%%%%%%%%%%%%%%%%%%%%%%%%%%%%%%
 
\section{Simulation using real data} \label{sec:Simulation_on_real_data} 
This section evaluates the performance of the proposed restoration algorithms by conducting two experiments. In both cases, the targets were put underwater while varying the concentration of Maalox\footnote{Maalox is a commercially available antacid medicine that strongly affects scattering without inducing significant optical absorption.} to change the attenuation level (i.e., attenuation factor $\alpha$) of the environment.  The images were acquired in June 2016 in the laboratory at Heriot-Watt University,  using a time-of-flight scanning sensor, based on TCSPC.  The transceiver system and data acquisition hardware used for this work are broadly similar to that described in \cite{MaccaroneOE2015}. The overall system had a jitter of $\approx 60$ps full width at half-maximum (FWHM) while we describe the other main parameters in Table \ref{tab:Parameters}. The section is divided into three main parts. The first part highlights the reconstruction of the reflectivity obtained in the highly attenuating environment.  The second part evaluates the restoration performance of the proposed algorithms while varying $\alpha$. The third part studies the restoration limits of the proposed algorithms while varying both $\alpha$ and the acquisition time  per pixel $t_{\textrm{acq}}$. 
\begin{table}[h] \centering
\centering \caption{Measurement parameters.}
\begin{tabular}{|c|c|}
\hline     \multirow{2}{*}{Laser system}   & Supercontinuum  \\
                                           &   laser system  \\
\hline     \multirow{1}{*}{Illum. Wavelength}   & $690$nm \\
\hline     \multirow{1}{*}{Laser Repetition Rate}   & $19.5$MHz \\
\hline     \multirow{1}{*}{Histogram bin width}   & $2$ps \\
\hline     \multicolumn{2}{c}{} \\ 
\hline     \multirow{2}{*}{Target 1}           & 2 reference targets       \\
                                               & with reflectivity 99$\%$  \\ 
                                               & and 10$\%$ (see Fig. \ref{fig:Exp1})   \\ 
\hline     \multirow{1}{*}{Scanned area}       & $5\times5$cm \\
\hline     \multirow{1}{*}{Number of pixels}   & $150\times150$ \\
\hline     \multirow{2}{*}{Acquisition time}   & Per pixel: $10$ms \\
                                               & Total: $\approx 4$ minutes \\
\hline     \multirow{1}{*}{Histogram length}   & $500$bins (after gating) \\ 
\hline     \multirow{1}{*}{Average optical power} & $\approx 670$nW \\ 
\hline     \multicolumn{2}{c}{} \\
\hline     \multirow{2}{*}{Target 2}   &  Pipe ($\approx 8\times5\times3.5$cm)   \\
                                     &  (see Fig. \ref{fig:Experiment2_Pipe_v2})   \\ 
\hline     \multirow{1}{*}{Scanned area}       & $5\times5$cm \\
\hline     \multirow{1}{*}{Number of pixels}   & $120\times120$ \\
\hline     \multirow{2}{*}{Acquisition time}   & Per pixel: $100$ms \\
                                               & Total: $\approx 24$ minutes \\
\hline     \multirow{1}{*}{Histogram length}   & $300$bins (after gating) \\ 
\hline     \multirow{1}{*}{Average optical power} & see Table \ref{tab:Exp_Pipe} \\ 
\hline  
\end{tabular}
\label{tab:Parameters}
\end{table}   
 
%%%%%%%%%%%%%%%%%%%%%%%%%%%%%%%%%%%%%%%%%%%%%%%%%%%
%%%%%%%%%%%%%%%%%%%%%%%%%%%%%%%%%%%%%%%%%%%%%%%%%%%
\subsection{Restoration of the reflectivity level} \label{subsec:Restoration_of_the_reflectivity_level}

\begin{figure}[h!]
\centering \subfigure[]{\includegraphics[width=0.58\figwidth,height=4cm]{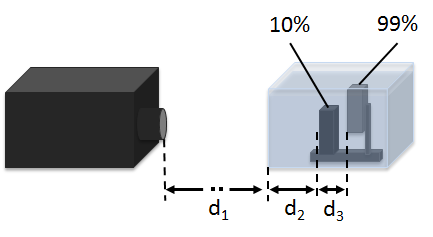}}
\subfigure[]{\includegraphics[width=0.38\figwidth,height=4cm]{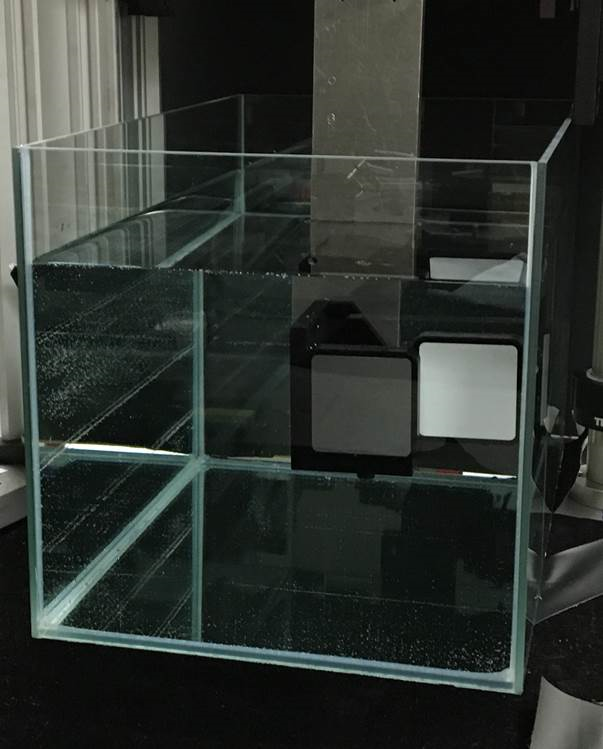}}
\caption{(a) Scheme of the first experiment with $d_1=1.57$m, $d_2=9.1$cm and $d_3=5.1$cm. (b) The two Spectralon  targets. } \label{fig:Exp1}
\end{figure}

\begin{figure}[h!]
\centering
\includegraphics[width=0.8\figwidth]{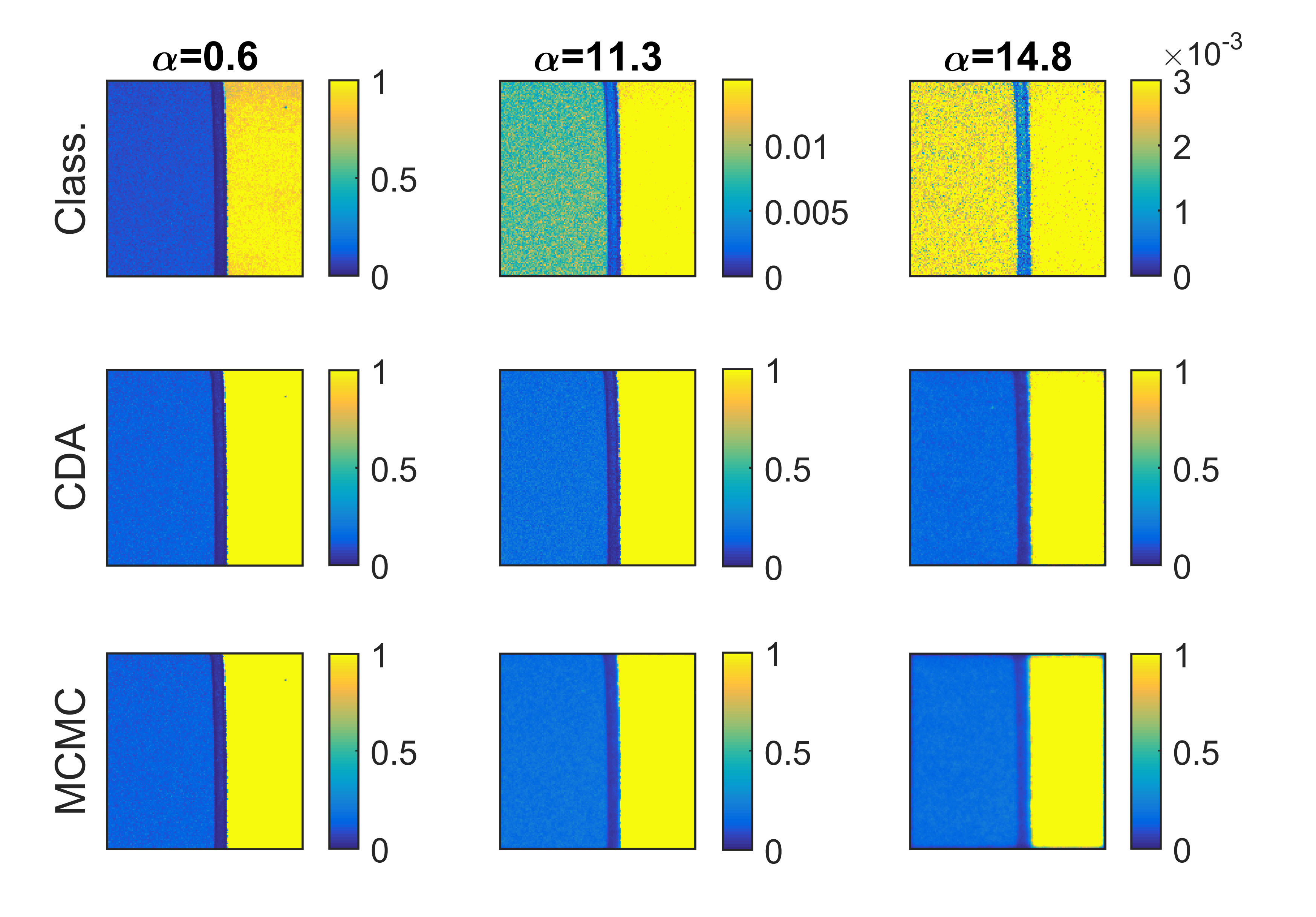}
\caption{Reflectivity images ($150\times 150$ pixels) obtained for $\alpha \in [0.6, 11.3, 14.8]$. (top)  classical XCorr approach, (Middle) proposed CDA algorithm, (Bottom) proposed MCMC algorithm.} \label{fig:Results_2Spec_Intens_Alphas}
\end{figure}

\begin{figure}[h!]
\centering
\includegraphics[width=0.8\figwidth]{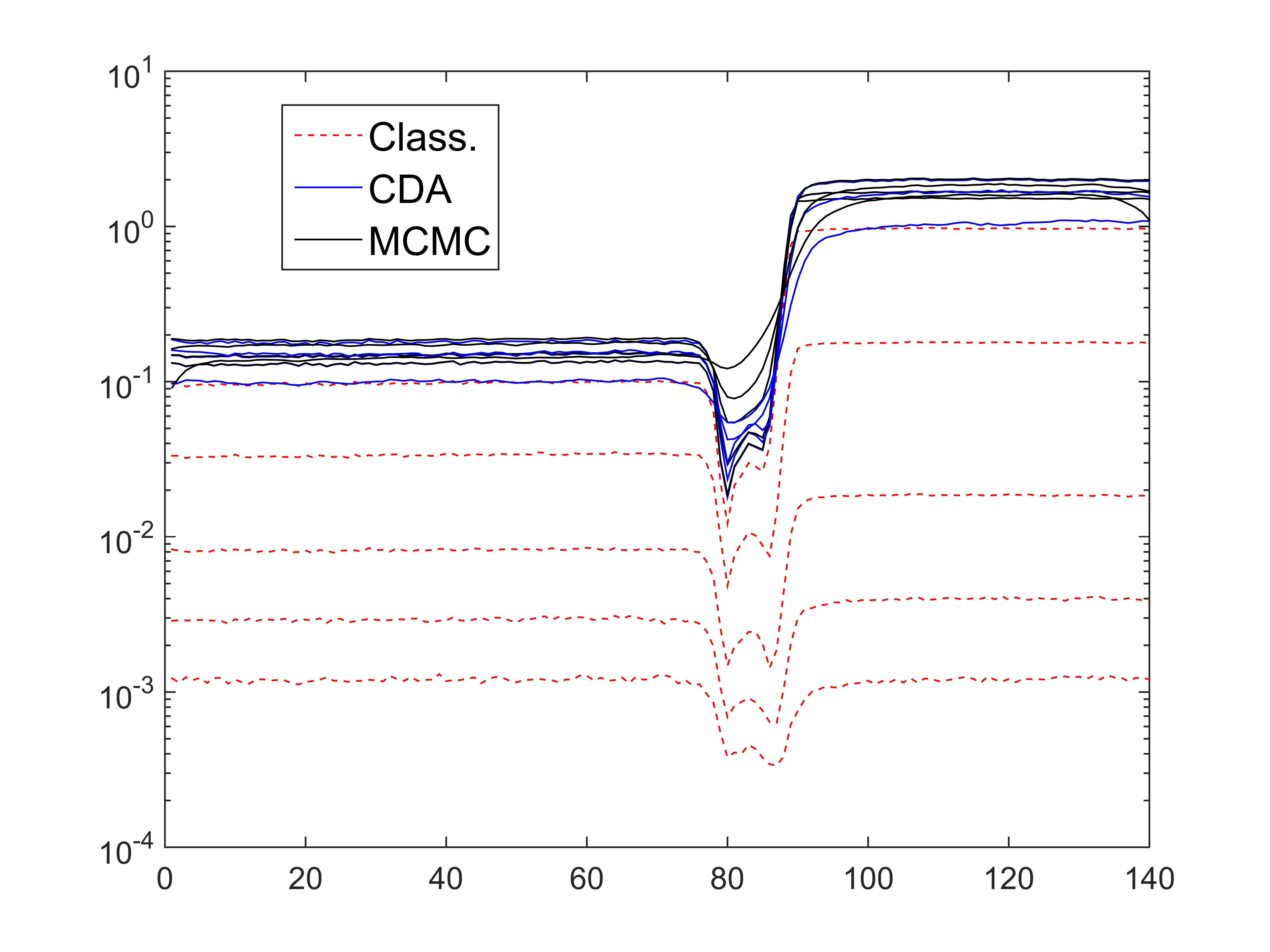}
\caption{Reflectivity lines ($150$ pixels) obtained for $\alpha \in [0.6, 5.2, 11.3, 14.8, 17.3]$ with the the classical XCorr approach (in dashed red lines), proposed CDA algorithm (in continuous blue lines) and the proposed MCMC algorithm (in continuous black lines).} \label{fig:Results_2Spec_Intens_Alphas_Lines}
\end{figure}

It is clear from \eqref{eqt:Observation_model} that if two objects are located in a attenuating environment (defined by $\alpha$) at a different distance from the sensor, they will be attenuated differently. This leads to the reflectivity distortion effect that is highlighted in this section. The experiment considers two reference targets (spectralon panels) with known reflectance (10 $\%$ and 99 $\%$),  that are put inside a tank of water (dim. $40\times 25\times 25$cm). The $99 \%$ reflectance spectralon panel is located at a longer distance from the sensor than the one at $10 \%$,  as shown in Fig. \ref{fig:Exp1}. Five data cubes (with $150\times 150$ pixels and $500$ time bins) were acquired for different attenuation levels  $\alpha \in [0.6, 5.2, 11.3, 14.8, 17.3]$ (obtained by varying the amount of Maalox in water).
Fig. \ref{fig:Results_2Spec_Intens_Alphas} shows the reflectivity images estimated by the classical  and the proposed algorithms. For clear water $\alpha=0.6$, the images show two levels of reflectivity related to the two spectralon panels, and separated by the edge of the spectralon which appears as blue vertical columns in the reflectivity maps. However, as $\alpha$ increases,  the reflectivity levels of the classical algorithm decrease differently in the two regions, until we obtain a uniform reflectivity map  (same level in the two regions) for $\alpha=14.8$. Indeed, the return from the 99 $\%$ reflectance spectralon panel is attenuated more than the  10 $\%$ reflectance one, since it is located at a longer distance. This distortion effect is corrected by the proposed CDA and MCMC algorithms that recover the true reflectivity level under the different  conditions of attenuation, as shown in Fig. \ref{fig:Results_2Spec_Intens_Alphas} (middle) and (bottom). Fig. \ref{fig:Results_2Spec_Intens_Alphas_Lines} shows the average of the rows of the reflectivity maps when varying  $\alpha$,  for the three algorithms. When increasing  $\alpha$, the classical algorithm (red lines) presents decreasing levels that end-up to be the same for $\alpha=14.8$ and slightly inversed for $\alpha=17.3$. The CDA and MCMC algorithms provide almost the same reflectivity results under different levels of $\alpha$. The observed small differences are mainly due to the presence of a high background noise for large $\alpha$, which affects the restoration performance of the proposed algorithms.   

%To do in the next experiment
%\begin{itemize}
%\item Measure the exact distances between the sensor-tank, tank-target1, target1-target2
%\item Take a picture of the experiment with clear water
%\item Measure the alpha for each quantity of Malox
%\item If possible: buy a new Spectralon 5$\%$ and tank
%\item Evaluate the effect of the focus ??
%\end{itemize}

~~\\
%\clearpage
%\clearpage
%
%\begin{figure*}
%\centering
%\includegraphics[width=1.95\figwidth]{images/Results_2Spec_Intens_Alphas2.png}
%\caption{Reflectivity maps ($150\times 150$ pixels) obtained for $\alpha \in[1, 5.5, 11.5, 15.5, 17.6]$. (top)  classical XCorr approach, (Middle) proposed CDA algorithm, (Bottom) proposed MCMC algorithm.} \label{fig:Results_2Spec_Intens_Alphas2}
%\end{figure*} 
%\begin{figure*}
%\centering
%\includegraphics[width=1.95\figwidth]{images/Results_2Spec_Depths_Alphas.png}
%\caption{Depth maps ($150\times 150$ pixels) obtained for $\alpha \in [1, 5.5, 11.5, 15.5, 17.6]$. (top)  classical XCorr approach, (Middle) proposed CDA algorithm, (Bottom) proposed MCMC algorithm.} \label{fig:Results_2Spec_Depths_Alphas}
%\end{figure*}
%
%\clearpage
%
%\newpage 
%\newpage 
%\newpage 

%%%%%%%%%%%%%%%%%%%%%%%%%%%%%%%%%%%%%%%%%%%%%%%%%%%
\subsection{Restoration of underwater depth and reflectivity images} \label{subsec:Restoration_of_UW_depth_and_reflectivity_images} 

\begin{figure}[h!] 
\centering \includegraphics[width=0.95\figwidth,height=4cm]{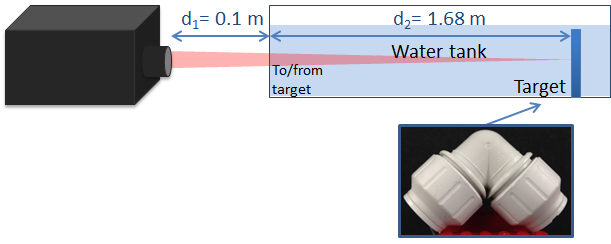}
\caption{Scheme of the second experiment showing a photograph of the plastic pipe target. } \label{fig:Experiment2_Pipe_v2}
\end{figure}

This section evaluates the performance of the proposed restoration algorithms when considering six real data cubes (of size $120 \times 120$ pixels and  $300$ time bins) of a plastic pipe, put at a stand-off distance of $1.68$m in water.  Fig. \ref{fig:Experiment2_Pipe_v2} presents the experimental scheme and shows a picture of the plastic pipe target.   The scans were performed with an acquisition time of $100$ms per pixel and different attenuation levels as shown in Table \ref{tab:Exp_Pipe}. The latter also shows the SBR and SNR levels estimated experimentally using a spectralon with known reflectivity. We provide these levels to link the analysis of this part to that on synthetic data.  
\renewcommand{\arraystretch}{1.2}
\begin{table}[h] \centering
\centering \caption{Attenuation levels for the underwater pipe measurement. The concentration of Maalox is obtained by dividing the volume of Maalox by the volume of the water ($67$ liters).}
\begin{tabular}{|c|c|c|c|c|c|c|}
  % after \\: \hline or \cline{col1-col2} \cline{col3-col4} ...
\hline     Fraction of Maalox &  \multirow{2}{*}{$ 0$}   & \multirow{2}{*}{$0.29$}   & \multirow{2}{*}{$ 0.60$}  & \multirow{2}{*}{$ 1$}  & \multirow{2}{*}{$1.22$}  & \multirow{2}{*}{$1.28$} \\
           ($\times 10^{-4}$) &   &    &   &    &   &  \\
\hline     AL                        &  $ 0.9$ & $ 2.5$   & $4.1$    & $ 6.7$ & $ 7.5$  & $8.1$\\
\hline     SBR                       &  $2322$ & $2576$   & $2344$   & $103$  & $13$    & $6$ \\  
\hline     SNR                       &  $505$  & $532$    & $592$    & $95$   & $32$    & $22$ \\ 
\hline    Average optical      &  \multirow{2}{*}{$0.5$}  & \multirow{2}{*}{$11$}     & \multirow{2}{*}{$235$}    & \multirow{2}{*}{$850$}  & \multirow{2}{*}{$850$}    & \multirow{2}{*}{$850$} \\
      power ($\mu$W)      &     &       &     &    &      &  \\
\hline 
\end{tabular}
\label{tab:Exp_Pipe}
\end{table}    
\begin{figure*}
\centering
\includegraphics[width=1.0\figwidth]{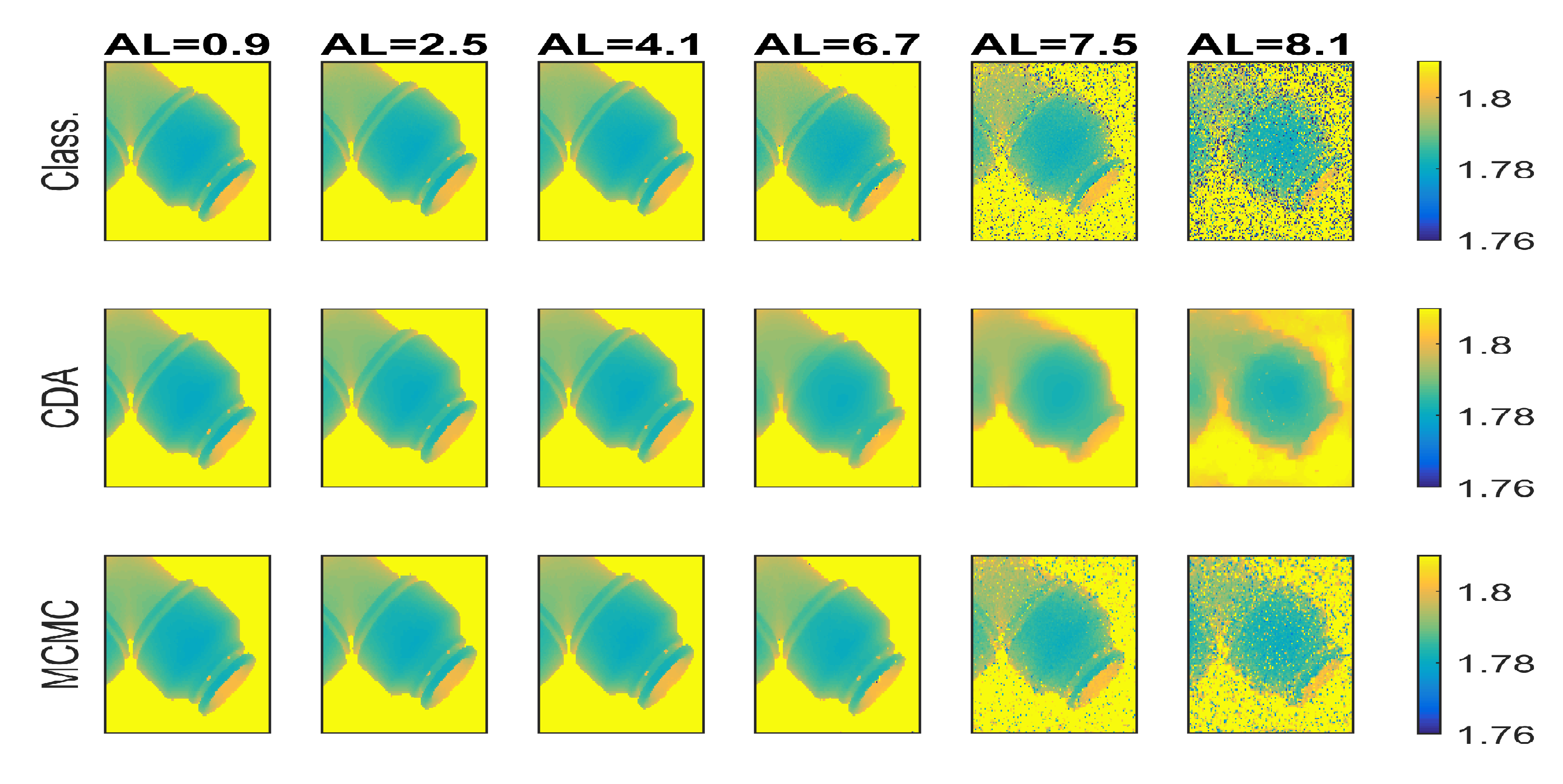}
\caption{Depth images ($120\times 120$ pixels) obtained for different attenuation factors with (top) the  classical XCorr approach, (middle)  the proposed CDA algorithm (bottom) and the proposed MCMC algorithm. The colormap is fixed for all images to [1.76,1.8] meters.} \label{fig:Results_Pipe_Depths_AL}
\end{figure*} 
\begin{figure*}
\centering
\includegraphics[width=1.0\figwidth]{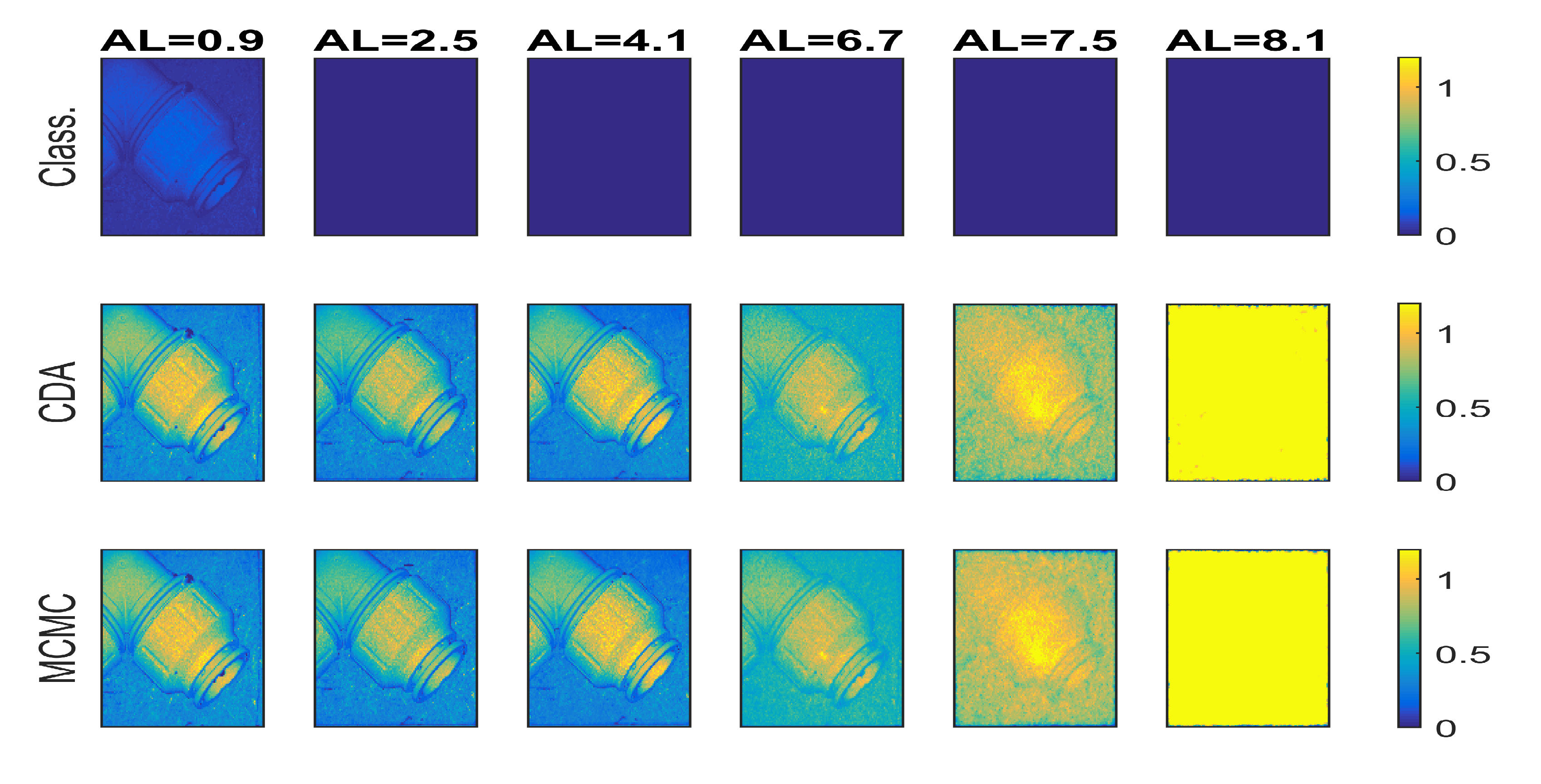}
\caption{Reflectivity images ($120\times 120$ pixels) obtained for different attenuation factors with (top) the classical XCorr approach, (middle)  the proposed CDA algorithm (bottom) and the proposed MCMC algorithm. The colormap is fixed for all images to [0, 1.2].} \label{fig:Results_Pipe_Intens_AL}
\end{figure*} 
  
Table \ref{tab:PSNR_Algos_AL} shows the SRE obtained with the algorithms. The algorithms proposed in this paper outperform the classical approach except for the reflectivity at the highest AL. In addition, the proposed algorithms show similar performance with  slightly better results for CDA whose hyperparameters have been adjusted to provide the highest SRE. Note that the MCMC algorithm also provide good results while automatically adjusting the MRF hyperparameters. However, this is achieved at the cost of significantly longer processing time, as highlighted in Table \ref{tab:Time_Algos}.  Figs. \ref{fig:Results_Pipe_Depths_AL} and \ref{fig:Results_Pipe_Intens_AL} show  examples of the obtained depth and reflectivity images with the algorithms for different ALs. The depths are restored well by the two algorithms while it can be seen that CDA over-smooths the pipe. The MCMC algorithm preserves more of the pipe contours while retaining some noise. These effects are mainly related to the estimated MRF hyperparameters that are different for the two algorithms. Considering the reflectivity images, the classical approach is largely affected by the environmental attenuation factor  while the proposed algorithms obtain acceptable results for $\textrm{AL} \leq 7.5$. For higher attenuation lengths, the restored reflectivity images are not satisfactory for two reasons: (i) the presence of a high background level and (ii) the measure of $\alpha$ is not too accurate because of the low signal level for these challenging scenarios, which affects the algorithms performance. 
\renewcommand{\arraystretch}{1.2}
\begin{table}  \centering
\centering \caption{SRE (in dB) of the restored depth and reflectivity images w.r.t. the attenuation lengths (AL).}
\begin{tabular}{|c|c|c|c|c|c|c|c|}
  \cline{3-8} \multicolumn{2}{c|}{}  &  \multicolumn{6}{c|}{Attenuation lengths} \\
	\cline{3-8} \multicolumn{2}{c|}{}  & 0.9     &  2.5   &  4.1   & 6.7     & 7.5      & 8.1     \\
\hline \multirow{3}{*}{Depth}        & Class.   &  $71.7$ & $49.9$ & $49.3$ & $54.8$   & $36.4$ & $34.2$ 
    \\
\cline{2-8}                          & CDA     & $82.2$ & $50.0$ & $49.4$ & $58.4$   & $50.3$ & $48.1$  \\ 
\cline{2-8}                          & MCMC    & $-$ & $50.0$ & $49.3$ & $56.1$   & $46.4$ & $43.2$  \\
\hline
\hline \multirow{3}{*}{Reflectivity}    & Class.   &  $1.5$ & $0.0$ & $0.0$ & $0.0$   & $0.0$ & $0.0$   \\
\cline{2-8}                          & CDA     & $59.5$ & $11.1$ & $11.0$ & $11.0$   & $3.4$ & $-7.5$  \\  
\cline{2-8}                          & MCMC    &  $-$ & $11.1$ & $11.0$ & $10.9$   & $2.7$ & $-8.4$  \\
  \hline
\end{tabular}
\label{tab:PSNR_Algos_AL}
\end{table} 
\renewcommand{\arraystretch}{1.2}
\begin{table}   \centering
\centering \caption{Processing time (in seconds).}
{\small\begin{tabular}{|c|c|c|c|c|c|c|}
  \cline{2-7} \multicolumn{1}{c|}{}  &  \multicolumn{6}{c|}{Attenuation lengths} \\
\cline{2-7} \multicolumn{1}{c|}{}    & 0.9    &  2.5   &  4.1   & 6.7     & 7.5      & 8.1    \\
\hline    \multicolumn{1}{|c|}{CDA}   & $ 21$ & $ 21$ & $ 21$ & $ 21$   & $ 18$ & $ 17$ \\ 
\hline   \multicolumn{1}{|c|}{MCMC}   & $529$ & $513$ & $514$ & $524$   & $496$ & $494$ \\
  \hline
\end{tabular}}
\label{tab:Time_Algos}
\end{table} 

%%%%%%%%%%%%%%%%%%%%%%%%%%%%%%%%%%%%%%%%%%%%%%%%%%%
\subsection{Performance w.r.t. the acquisition times and the attenuation factor} \label{subsec:Performance_wrt_the_acquisition_times_and_the_attenuation_factor} 
\begin{figure}
\centering
\includegraphics[width=0.8\figwidth]{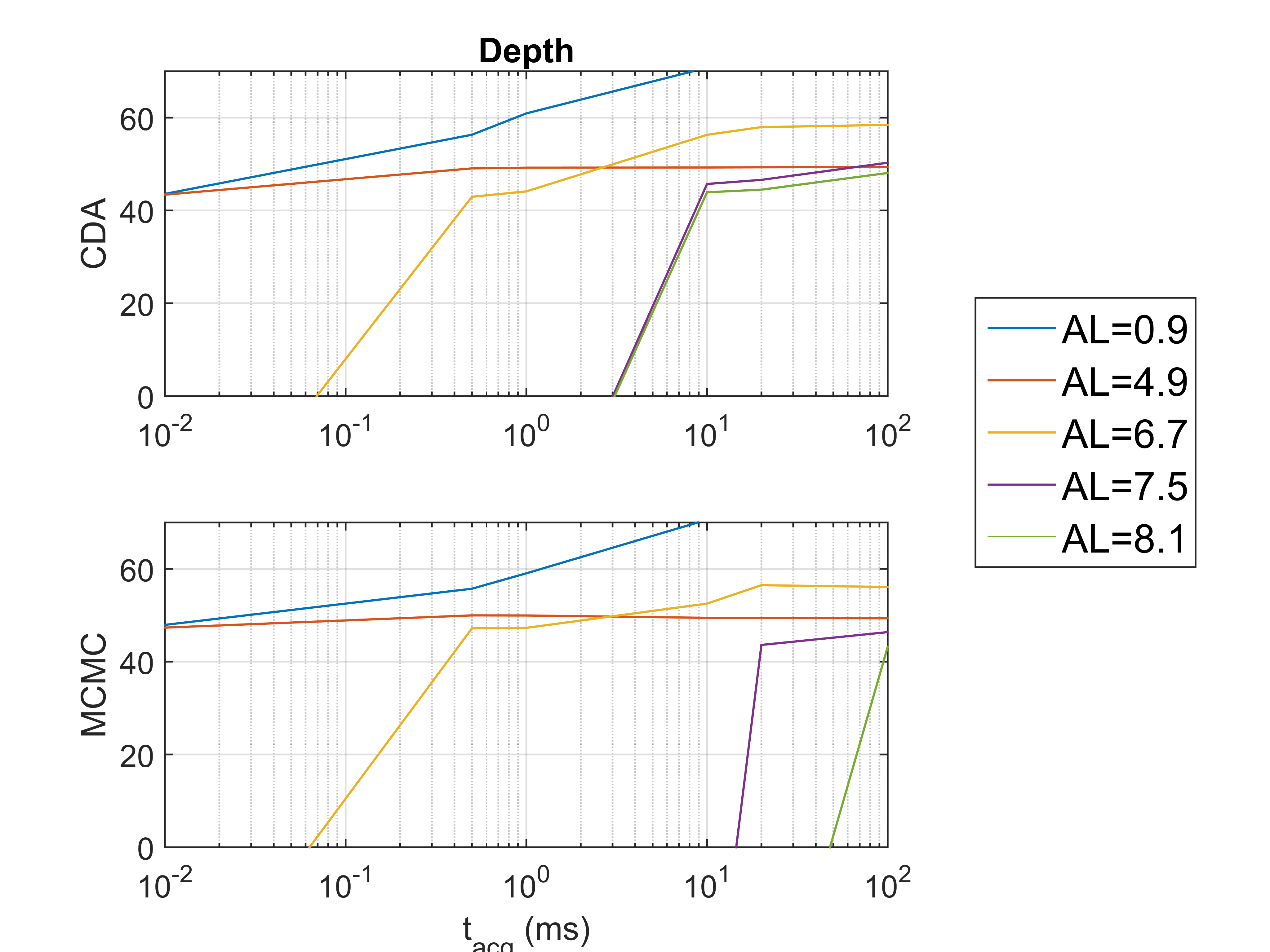}
\caption{Depth SRE obtained w.r.t. the acquisition time per-pixel ($t_{\textrm{acq}}$) for  different attenuation levels. (top) CDA, (bottom) MCMC.} \label{fig:Results_Pipe_Depth_AL_Time}
\end{figure}

\begin{figure}
\centering
\includegraphics[width=0.8\figwidth]{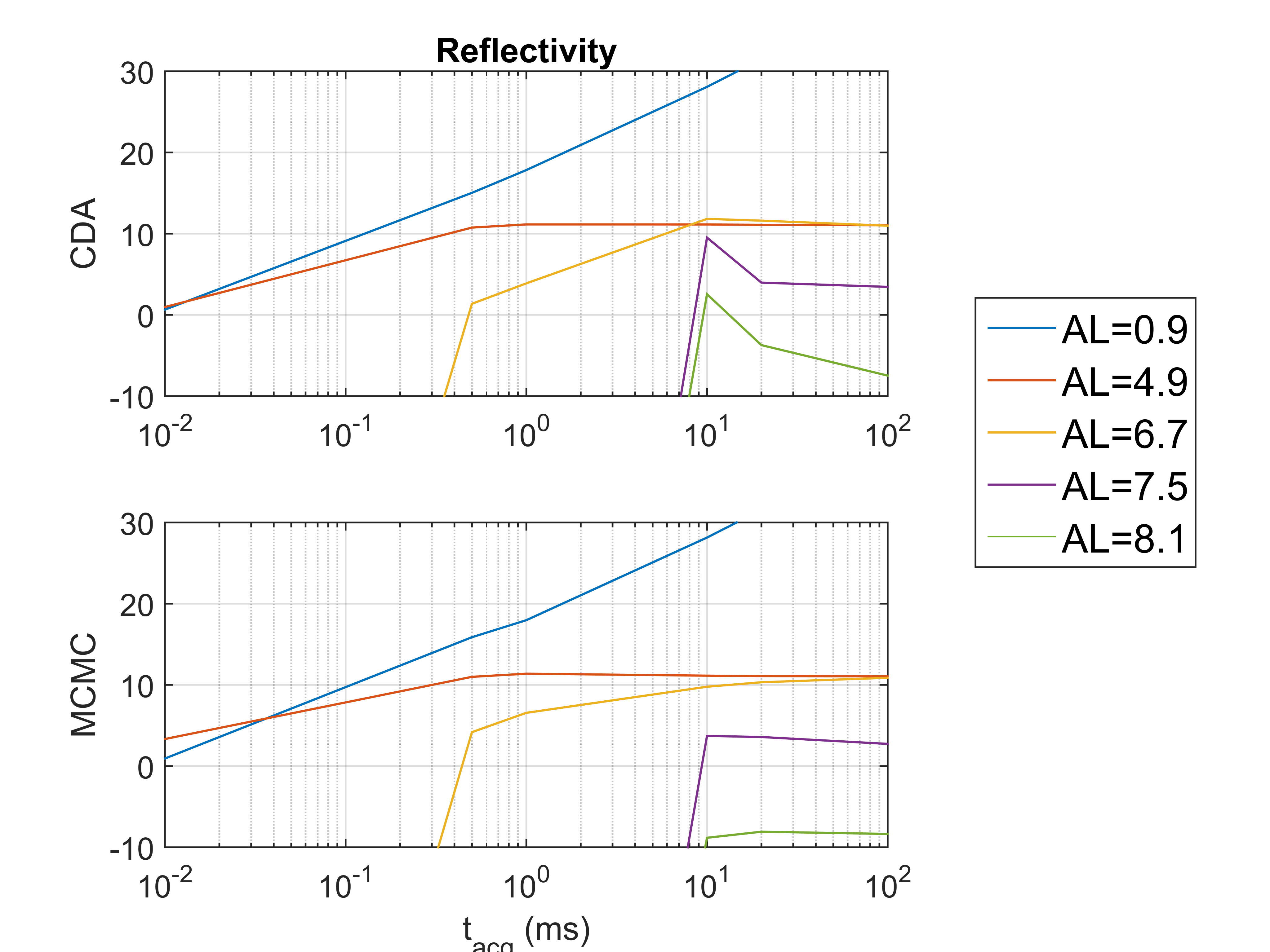}
\caption{Reflectivity SRE obtained w.r.t. the acquisition time per-pixel ($t_{\textrm{acq}}$) for  different attenuation levels. (top) CDA, (bottom) MCMC.} \label{fig:Results_Pipe_Intens_AL_Time}
\end{figure}

This section explores the performance of the proposed algorithms when dealing with a reduced number of photons due to a reduced acquisition time or an attenuating environment. This evaluation is important to state the possible level of attenuation that can be dealt with the proposed algorithms. In this experiment, we will consider the data used in the previous section with $t_{\textrm{acq}}=100$ms (see Fig. \ref{fig:Experiment2_Pipe_v2}). Note however that the data format of timed events allows the construction of photon timing histograms associated with shorter acquisition times, after measurement, as the system records the time of arrival of each detected photon. Here, we evaluate our algorithms for acquisition times ranging from $0.01$ms to $100$ms per pixel. Table \ref{tab:Tacq_AL_Algos} reports the percentage of non-empty pixels  w.r.t. $t_{\textrm{acq}}$ and AL. As expected, this percentage is higher for high $t_{\textrm{acq}}$ or low AL. 
Figs. \ref{fig:Results_Pipe_Depth_AL_Time} and \ref{fig:Results_Pipe_Intens_AL_Time} show the SRE as a function of  $t_{\textrm{acq}}$ for different attenuation lengths. First note that the MRF parameters of the CDA algorithm have been adjusted to provide the best SRE results, which explain why CDA outperforms MCMC in some cases. As expected, the algorithms performance generally decreases while reducing the acquisition times or increasing the attenuation levels. As AL increases, the algorithms require more acquisition time (i.e., more informative pixels) in order to obtain an acceptable performance. The latter are generally obtained for a percentage higher than $30 \%$ of non-empty pixels and $\textrm{AL} \leq 7.5$.
For example, when $\textrm{AL}=7.5$, the CDA algorithm requires that $t_{\textrm{acq}}>10$ms to reach a good performance both for depth and reflectivity.   
Therefore, given an attenuating environment defined by $\alpha$, these results allow the setting of the required acquisition times to obtain a given level of accuracy.   
 
\renewcommand{\arraystretch}{1.2}
\begin{table}   \centering
\centering \caption{Percentage of useful pixels w.r.t. $t_{\textrm{acq}}$ and AL.}
{\small\begin{tabular}{|c|c|c|c|c|c|c|c|}
  \cline{3-8} \multicolumn{2}{c|}{}  &  \multicolumn{6}{c|}{Attenuation lengths} \\
\cline{3-8} \multicolumn{2}{c|}{}    &  0.9    &  2.5   &  4.1   & 6.7     & 7.5      & 8.1    \\
\hline  &  $0.01$   & $32.1$  & $30.8$  & $35.6$ & $1.0$   & $0.3$ & $0.4$ \\ 
\cline{2-8}   &  $0.1$   & $91.9$  & $91.2$  & $92.1$ & $9.2$   & $3.3$ & $2.5$ \\
\cline{2-8} \multirow{2}{*}{$t_{\textrm{acq}}$}  & $0.5$   & $99.7$  & $99.7$  & $99.8$ & $34.2$   & $14.5$ & $11.8$   \\ 
\cline{2-8}   &  $1$     & $99.9$  & $100.0$ & $100.0$ & $51.5$   & $25.0$ & $20.7$     \\
\cline{2-8}   \multirow{2}{*}{$(\textrm{ms})$} & $2$     & $99.9$  & $100.0$ & $100.0$ & $67.1$   & $37.0$ & $32.0$     \\ 
\cline{2-8}   & $10$    & $100.0$ & $100.0$ & $100.0$ & $88.6$   & $53.5$ & $43.0$     \\
\cline{2-8}   &  $20$    & $100.0$ & $100.0$ & $100.0$ & $95.4$   & $61.7$ & $48.1$     \\ 
\cline{2-8}   & $100$   & $100.0$ & $100.0$ & $100.0$ & $100.0$   & $84.0$ & $72.6$\\
  \hline
\end{tabular}}
\label{tab:Tacq_AL_Algos}
\end{table}

%%%%%%%%%%%%%%%%%%%%%%%%%%%%%%%%%%%%%%%%%%%%%%%%%%%
%%%%%%%%%%%%%%%%%%%%%%%%%%%%%%%%%%%%%%%%%%%%%%%%%%%
%%%%%%%%%%%%%%%%%%%%%%%%%%%%%%%%%%%%%%%%%%%%%%%%%%% 
\section{Conclusions} \label{sec:Conclusions}   
This paper introduced a hierarchical Bayesian model and two estimation algorithms for the restoration of depth and reflectivity obtained in the limit of very low photon counts and significant attenuation. The algorithms were designed to provide the single-photon community with useful, relatively fast, and practical tools for the image restoration. Using some assumptions, a new formulation was introduced leading to a log-concave likelihood that is only expressed using preliminary estimates of the DR images. The restoration of these two images was achieved by considering two MRF based prior distributions ensuring spatial correlation between the pixels. The resulting joint posterior distribution was used to approximate the Bayesian estimators. First, a Markov chain Monte Carlo procedure based on a Metropolis-within-Gibbs algorithm was used to sample the posterior of interest and to approximate the MMSE estimators of the unknown parameters using the generated samples. Second, a coordinate descent approach  using an alternating direction method of multipliers algorithm was used to approximate the  maximum a posteriori estimators. Both algorithms showed comparable performance while providing different characteristics, i.e., the MCMC algorithm was fully automatic while the CDA algorithm required a reduced computational time. Results on both synthetic and real data showed the ability of the proposed algorithms to correct the reflectivity distortion effect, and to restore the depth and reflectivity images obtained in highly attenuating environments. 
Future work includes relaxing some of the assumptions of this paper, which might lead to better performance at the price of a higher computational cost. Generalizing the algorithms to account for target with multiple depth returns \cite{HernandezPAMI2007,Hernandez_MarinPAMI2008} is also an interesting issue which is worthy of investigation.

%%%%%%%%%%%%%%%%%%%%%%%%%%%%%%%%%%%%%%%%%%%%%%%%%%%%
%%%%%%%%%%%%%%%%%%%%%%%%%%%%%%%%%%%%%%%%%%%%%%%%%%%% 
%\section*{Acknowledgment}
%The authors acknowledge the support of the DSTL National PhD Scheme for financing A. Maccarone.
 %

\appendix[ADMM algorithm] \label{app:ADMM_algorithm}
Consider the optimization problem \vspace{-0.25cm}
\begin{equation}
\operatornamewithlimits{\textrm{argmin}}\limits_{\bst} \mathcal{C} \left(\bst \right)  = \operatornamewithlimits{\textrm{argmin}}\limits_{\bst} \sum_{j=1}^{J}{g_j \left(\bsH^{(j)} \bst\right)} \vspace{-0.25cm}
\label{eqt:Cost_FunADMM}
\end{equation} 
where $\bst \in \mathds{R}^{N\times 1}$,  $g_j: \mathds{R}^{p_j} \rightarrow \mathds{R}$ are closed, proper, convex functions, and $\bsH^{(j)} \in \mathds{R}^{p_j\times N}$ are arbitrary matrices. After denoting $\bsu^{(j)}=\bsH^{(j)} \bsz  \in \mathds{R}^{p_j}$ and introducing the auxiliary variable $\bsd^{(j)} \in\mathds{R}^{p_j}$,  the authors in \cite{BioucasDias2012,Figueiredo_TIP2010} introduced the ADMM variant summarized in Algo. \ref{alg:ADMM_for_depth_estimation} to solve \eqref{eqt:Cost_FunADMM}. This algorithm converges when the matrix $\bsM = \left[ \sum_{j=1}^{J}{ \left(\bsH^{(j)}\right)^{\top} \bsH^{(j)} }  \right]$ has full rank, and the optimization problems in line $10$ are solved exactly or if their sequences of errors are absolutely summable \cite{Figueiredo_TIP2010}. In our case, we have
\begin{eqnarray}
g_1\left(u^{(1)}_{i,j}\right) =  \frac{\left(u^{(1)}_{i,j} - t_{i,j}^{\textrm{ML0}} + \alpha \sigma^2  \right)^2}{\frac{2 \sigma^2}{c_2 r_{i,j}^{\textrm{ML0}} } }  + c_2 r_{i,j}   \exp^{\left(-\alpha u^{(1)}_{i,j}\right)}, \nonumber \\
g_2\left(\bsu^{(2)}\right) = \eta ||\bsu^{(2)}||_1, \textrm{ and  }\;\;  g_3\left(\bsu^{(3)}\right) = \textit{i}_{\mathds{R}_+}\left(\bsu^{(3)}\right), \;\;\; \; 
\label{eqt:Def_gi_Hi_ADMMTV}
\end{eqnarray}
where $\bsH^{(1)}=\bsK$  is a $Q\times N$ binary matrix that contains a single non-zero value (equals to 1) on each line to model the loss of some image pixels and $Q$ is the number of non-empty pixels, $\bsH^{(2)}$ denotes the TV linear operator as described in \cite{BioucasDias2012}, and $\bsH^{(3)}=\mathds{I}_{N}$.  These matrices lead to $\bsM = \mathds{I}_{N} + \bsK^{\top}\bsK + \bsH^{(2)\top} \bsH^{(2)}$  which is a full rank matrix ($\bsK^{\top}\bsK$ is a diagonal matrix whose values equal 0 in the position of missing pixels and 1 otherwise).    
The updates of $\bsu^{(2)}, \bsu^{(3)}$ in line  $10$ of Algo. \ref{alg:ADMM_for_depth_estimation} are straightforward and lead to exact solutions. For $\bsu^{(1)}$, the optimization problem has been solved using few iterations of the Newton method \cite{Bertsekas1995}. 
Regarding the solution of \eqref{eqt:Cost_FunADMM}, note that  $g_1+g_2+\textrm{TV}$  is proper, coercive, lower semi-continuous, and  strictly convex for $r_{i,j}^{\textrm{ML0}}>0$, and $r_{i,j}>0$ (which is satisfied). Since $\bsK$ is injective,  we obtain that $\mathcal{C}(\bst) = g_1  (\bsK \bst)  + g_2(\bst) + \eta \textrm{TV}(\bst)$ is proper, coercive, lower semi-continuous, and  strictly convex, thus, there is a unique minimizer for $\mathcal{C}(\bst)$ (see for example \cite{Bertsekas1995,CombettesMMS2005,Figueiredo_TIP2010}).  
%%% quad:  str. convex (for rML0>0), proper, coercive, lsc
%%% expo:  str. convex (for r_ij>0), proper,  lsc
%%% quad+expo+iR+: str. convex (for rML0>0,r_ij>0), proper, coercive, lsc
%%% TV:  convex, proper,  lsc
%%% quad+expo+iR+: str. convex (for rML0>0,r_ij>0), proper, coercive, lsc 
The authors invite the reader to consult \cite{Figueiredo_TIP2010,Afonso_TIP2011,BioucasDias2012} for more details regarding the ADMM algorithm and its convergence characteristics.
\begin{algorithm}
\caption{ADMM for depth estimation} \label{alg:ADMM_for_depth_estimation}
\begin{algorithmic}[1]
       \STATE \underline{Initialization}
       \STATE Initialize $\bsu^{(j)}_0, \bsd^{(j)}_0, \forall j$, $\mu$. Set $k\leftarrow 0$, conv$\leftarrow 0$    
       \WHILE{conv$=0$}
                  \FOR{j=1:J} 
							             \STATE $\xi^{(j)}_{k} \leftarrow \bsu^{(j)}_{k}+\bsd^{(j)}_{k}$,
							       \ENDFOR
							 \STATE $\bst_{k+1} \leftarrow \bsM^{-1}   \sum_{j=1}^{J}{\left(\bsH^{(j)}\right)^{\top} \xi^{(j)}_{k}}$,							
							   \FOR{j=1:J} 
							             \STATE $\bsv^{(j)}_{k} \leftarrow \bsH^{(j)} \bst_{k+1} - \bsd^{(j)}_{k}$,
													 \STATE $\bsu^{(j)}_{k+1} \leftarrow  \operatornamewithlimits{\textrm{argmin}}\limits_{\bsm}  \frac{\mu}{2}   ||\bsm - \bsv^{(j)}_{k}||^2 + g_j\left(\bsm \right)$,
							   \ENDFOR 
						     \FOR{j=1:J} 
							             \STATE $\bsd^{(j)}_{k+1}  \leftarrow \bsd^{(j)}_{k} - \left( \bsH^{(j)} \bst_{k+1} - \bsu^{(j)}_{k+1}\right)$,
							   \ENDFOR 
               \STATE $k = k  + 1$
       \ENDWHILE 
\end{algorithmic}
\end{algorithm}

%%%%%%%%%%%%%%%%%%%%%%%%%%%%%%%%%%%%%%%%%%%%%%%%%%%
%\renewcommand{\baselinestretch}{0.85}
%\footnotesize
%\small
\bibliographystyle{IEEEtran}
\bibliography{biblio_all}

\end{document}

%% file: notations.tex
\input com_notations.tex

\newcounter{algo}
\renewcommand{\thealgo}{\arabic{algo}}

%% file: com_notations.tex
\def\bsb{{\boldsymbol{b}}}

\def\bsd{{\boldsymbol{d}}}

\def\bsm{{\boldsymbol{m}}}

\def\bsr{{\boldsymbol{r}}}

\def\bst{{\boldsymbol{t}}}
\def\bsu{{\boldsymbol{u}}}
\def\bsv{{\boldsymbol{v}}}
\def\bsw{{\boldsymbol{w}}}
\def\bsx{{\boldsymbol{x}}}
\def\bsy{{\boldsymbol{y}}}
\def\bsz{{\boldsymbol{z}}}

\def\bsH{{\boldsymbol{H}}}

\def\bsK{{\boldsymbol{K}}}

\def\bsM{{\boldsymbol{M}}}

\def\bsY{{\boldsymbol{Y}}}

\def\calG{{\mathcal{G}}}

\def\calI{{\mathcal{I}}}